\documentclass[pra,aps,twocolumn,amsmath,amssymb]{revtex4-1}

\usepackage{graphicx}
\usepackage{MnSymbol}
\usepackage{bbold} 

\begin{document}

\title{Radiation pressure on single atoms: generalization of an exact analytical approach to multilevel atoms}

\author{L. Podlecki}
\affiliation{Institut de Physique Nucl\'eaire, Atomique et de Spectroscopie, \\ CESAM, University of Liege, B\^at.~B15, Sart Tilman, Li\`ege 4000, Belgium}

\author{J. Martin}
\affiliation{Institut de Physique Nucl\'eaire, Atomique et de Spectroscopie, \\ CESAM, University of Liege, B\^at.~B15, Sart Tilman, Li\`ege 4000, Belgium}

\author{T. Bastin}
\affiliation{Institut de Physique Nucl\'eaire, Atomique et de Spectroscopie, \\ CESAM, University of Liege, B\^at.~B15, Sart Tilman, Li\`ege 4000, Belgium}

\date{June 1, 2021}

\begin{abstract}
In a recent work, we provided a standardized and exact analytical formalism for computing in the semiclassical regime the radiation force experienced by a two-level atom interacting with any number of plane waves with arbitrary intensities, frequencies, phases, and propagation directions [J. Opt. Soc. Am. B \textbf{35}, 127-132 (2018)]. Here, we extend this treatment to the multilevel atom case, where degeneracy of the atomic levels is considered and polarization of light enters into play. A matrix formalism is developed to this aim.
\end{abstract}

\def\contFracOpe{%
    \operatornamewithlimits{%
        \mathchoice{
            \vcenter{\hbox{\huge $\mathcal{K}$}}%
        }{
            \vcenter{\hbox{\Large $\mathcal{K}$}}%
        }{
            \mathrm{\mathcal{K}}%
        }{
            \mathrm{\mathcal{K}}%
        }
    }
}

\maketitle

\section{Introduction}

The mechanical action of laser light on atoms led to a great number of spectacular experiments and achievements in atomic physics during the past decades (see, e.g., Refs.~\cite{Schieder1972,Phillips1982,Chu1986,Kasevich1989,Timp1992,Anderson1995,Chabe2008,Corder2015}). In the semiclassical regime, where the atomic motion is treated classically, the resonant laser radiation produces a mechanical force on the atomic center-of-mass. Recently, we provided a standardized exact analytical treatment of the mechanical action induced by an arbitrary set of plane waves on a two-level atom~\cite{Podlecki2018}. In particular, we showed that the light force always reaches a periodic regime shortly after establishment of the interaction and we provided an exact yet simple expression of all related Fourier components of the force in this regime. The mean net force $\mathbf{F}$ has been shown to be expressible in all cases (coherent and incoherent) in the form $\mathbf{F} = \sum_{j=1}^N \mathbf{F}_j$, with
\begin{equation} \label{F_KnownExpression}
\mathbf{F}_j = \frac{\Gamma}{2} \frac{s_j}{1+s} \hbar \mathbf{k}_j
\end{equation}
the force exerted by the $j$th plane wave in presence of all other waves. Here, $N$ is the number of plane waves lightening the atom, $\Gamma$ is the spontaneous de-excitation rate of the upper level of the transition, $\hbar \mathbf{k}_j$ is the $j$th plane wave photon momentum, and $s=\sum_j s_j$, with $s_j$ a \emph{generalized} saturation parameter
\begin{equation} \label{sj_KnownExpression}
s_j = \textrm{Re} \left[ \frac{\Omega_j}{\Gamma/2 - i \delta_j} \sum\limits_{l=1}^N \frac{\Omega_l^*}{\Gamma} q_{m_{lj}} \right],
\end{equation}
where $\Omega_j$ and $\delta_j$ are the Rabi frequency and the detuning of plane wave $j$, and $q_{m_{lj}}$ are complex numbers obtained from the solution of an infinite system of equations~\cite{Podlecki2018}. In the low-intensity and incoherent regime, we showed that Eq.~(\ref{sj_KnownExpression}) simplifies to the standard expression of the saturation parameter, i.e.,
\begin{equation} \label{sj_StandardExpression}
s_j = \frac{\vert \Omega_j \vert^2/2}{\Gamma^2/4 + \delta_j^2} .
\end{equation}

One then might naturally bring up the question of how this formalism extends to the multilevel atom case, where Zeeman sublevels and arbitrary polarization of light are considered. In this case, a closed form of the mean net force $\mathbf{F}$ is only known in specific situations. For instance, for an atom with degenerate ground and excited states of angular momenta $J_g$ and $J_e$, respectively, the stationary force exerted by a linearly polarized plane wave is either 0 (if $\Delta J \equiv J_e - J_g = -1$, or $\Delta J = 0$ with integer $J_g$) or reads
\begin{equation} \label{F_Gao}
\mathbf{F} = \frac{\Gamma}{2} \frac{s}{b+s} \hbar \mathbf{k} ,
\end{equation}
with $\hbar \mathbf{k}$ the plane wave photon momentum, $s$ the standard saturation parameter~(\ref{sj_StandardExpression}) and $b$ a parameter depending on $J_g$ and $J_e$~\cite{Gao1993}. Other specific cases have also been studied in the lin$\perp$lin and $\sigma^+$-$\sigma^-$ configurations~\cite{Chang1999,Chang2001Comp,Chang2001,Chang2002,Chuang2009}. However, no exact analytical extension of Eq.~(\ref{F_Gao}) is known for general atomic and laser configurations. In practice, purely numerical approaches are enforced in this case (see, e.g., Refs.~\cite{Castin1989,Molmer1991B,Molmer1991,Paspalakis2000,Sukharev2010,Rutherford2010,Xuereb2010}).

Here, we extend the formalism developed in Ref.~\cite{Podlecki2018} to the multilevel atom case and we show that an exact analytical expression of the mechanical force exerted by a set of plane waves with arbitrary intrinsic properties (intensity, phase, frequency, polarization, and propagation direction) can be similarly obtained after a transient regime. A matrix formalism is developed to this aim. The paper is organized as follows. In Section~\ref{SecForce}, the generalized optical Bloch equation (OBE) formalism is developed and we extract the sought exact expression of the radiation pressure force in the most general configuration. In Section~\ref{SecSpecific}, specific regimes where interesting simplifications occur are investigated and we draw conclusions in Section~\ref{SecCcl}. Finally, two Appendices close this paper, where we detail the effect of a reference frame rotation on the OBEs (Appendix~\ref{AppendixA}) and explicit values of specific matrices are given (Appendix~\ref{AppendixB}).

\section{Model and radiation pressure force} \label{SecForce}

\subsection{Hamiltonian and master equation}

We consider an atom with two degenerate levels of energy $E_e \equiv \hbar \omega_e$ and $E_g \equiv \hbar \omega_g$ ($E_e > E_g$), and of total angular momenta $J_e$ and $J_g$, respectively. The Zeeman sublevels are denoted $\vert J_e , m_e \rangle$ and $\vert J_g, m_g \rangle$. We consider an electric dipole transition ($\Delta J \equiv J_e - J_g \in \{ 0 , \pm 1 \}$) with angular frequency $\omega_e - \omega_g \equiv \omega_{eg}$. The atom interacts with a classical electromagnetic field $\mathbf{E} (\mathbf{r}, t)$ resulting from the superposition of $N$ arbitrary plane waves: $\mathbf{E} (\mathbf{r}, t) = \sum_{j=1}^N \mathbf{E}_j (\mathbf{r}, t)$, with $\mathbf{E}_j (\mathbf{r}, t) = (\mathbf{E}_j/2) e^{i (\omega_j t - \mathbf{k}_j \cdot \mathbf{r} + \varphi_j)} + \textrm{c.c.}$ Here, $\omega_j$, $\mathbf{k}_j$, and $\varphi_j$ are the angular frequency, the wave vector, and the phase of the $j$th plane wave, respectively, and $\mathbf{E}_j \equiv E_j \boldsymbol{\epsilon}_j$, with $E_j > 0$ and $\boldsymbol{\epsilon}_j = \sum_q \epsilon_{j, q} \mathbf{e}^q$ the normalized polarisation vector of the corresponding wave written in the upper-index spherical basis $\{ \mathbf{e}^q , q = 0, \pm 1 \}$~\cite{footnote}. Nonzero $\epsilon_{j, 0}$ and $\epsilon_{j, \pm 1}$ components correspond to so-called $\pi$ and $\sigma^\pm$ polarization components of radiation, respectively. Accordingly, the vectors $\mathbf{e}^0$, $\mathbf{e}^{\pm 1}$ are also denoted by $\boldsymbol{\pi}$, $\boldsymbol{\sigma}^\pm$, respectively. As in our analysis of the two-level case~\cite{Podlecki2018}, the quasi-resonance condition is fulfilled for each plane wave ($\vert \delta_j \vert \ll \omega_{eg}, \forall j$, where $\delta_j = \omega_j - \omega_{eg}$ is the detuning). We also define a weighted mean frequency $\overline{\omega} = \sum_j \kappa_j \omega_j$ of the plane waves together with a weighted mean detuning $\overline{\delta} = \sum_j \kappa_j \delta_j = \overline{\omega} - \omega_{eg}$, with $\{ \kappa_j \}$ an \textit{a priori} arbitrary set of weighting factors ($\kappa_j \ge 0$ and $\sum_j \kappa_j = 1$).

The time evolution of the atomic density operator $\hat{\rho}$ is governed by the standard master equation~\cite{Molmer1993}
\begin{equation} \label{MasterEqMolmer}
\frac{d}{dt} \hat{\rho} (t) = \frac{1}{i \hbar} [ \hat{H} (t) , \hat{\rho} (t) ] + \mathcal{D} (\hat{\rho} (t))
\end{equation}
in which $\hat{H} (t) = \hbar \omega_e \hat{P}_e + \hbar \omega_g \hat{P}_g - \hat{\mathbf{D}} \cdot \mathbf{E} (\mathbf{r}, t)$ and
\begin{equation}
\mathcal{D} ( \hat{\rho} ) = - (\Gamma/2) ( \hat{P}_e \hat{\rho} + \hat{\rho} \hat{P}_e ) + \Gamma \sum_q ( {\mathbf{e}^q}^* \cdot \hat{\mathbf{S}}^- ) \hat{\rho} ( \mathbf{e}^q \cdot \hat{\mathbf{S}}^+ ) .
\end{equation}
Here, $\hat{P}_k \equiv \sum_{m_k} \vert J_k, m_k \rangle \langle J_k, m_k \vert$ $(k=e,g)$, $\hat{\mathbf{D}}$ is the atomic electric dipole operator, $\mathbf{r}$ is the atom position in the electric field, $\Gamma$ is the spontaneous de-excitation rate of each upper sublevel, and the $\hat{\mathbf{S}}^\pm$ operators are defined according to $\mathbf{e}^q \cdot \hat{\mathbf{S}}^+ \vert J_g, m_g \rangle = C_{m_g}^{(q)} \vert J_e , m_g + q \rangle$, $\mathbf{e}^q \cdot \hat{\mathbf{S}}^+ \vert J_e, m_e \rangle = 0$, and ${\mathbf{e}^q}^* \cdot \hat{\mathbf{S}}^- = ( \mathbf{e}^q \cdot \hat{\mathbf{S}}^+ )^\dagger$, with
\begin{equation}
\mathcal{C}_m^{(q)} \equiv \langle J_g, m ; 1, q \vert J_e, m + q \rangle ,
\end{equation}
where $\langle j_1, m_1 ; j_2, m_2 \vert j, m \rangle$ is the Clebsch-Gordan coefficient corresponding to the coupling of $\vert j_1, m_1 \rangle$ and $\vert j_2, m_2 \rangle$ into $\vert j, m \rangle$.

\subsection{Optical Bloch equations}

All matrix elements $\rho_{(J_k,m_k),(J_l,m_l)} \equiv \langle J_k, m_k \vert \hat{\rho} \vert J_l, m_l \rangle$ $(k, l = e, g)$ are dependent variables due to hermiticity and unit trace of $\hat{\rho}$. We consider here the column vector of real and independent variables $\mathbf{x} = ( \mathbf{x}_o^T , \mathbf{x}_\xi^T )^T$, with $\mathbf{x}_o$ a column vector of optical coherences and $\mathbf{x}_\xi \equiv (\mathbf{x}_p^T , \mathbf{x}_Z^T)^T$, where $\mathbf{x}_p$ is a column vector of populations and $\mathbf{x}_Z$ a column vector of Zeeman coherences. We defined
\begin{equation} \label{cohOpt}
\begin{split}
\mathbf{x}_o = \begin{pmatrix} \mathbf{x}_o^{(-(J_e+J_g))} \\ \vdots \\ \mathbf{x}_o^{(J_e+J_g)} \end{pmatrix} , \textrm{ with } \mathbf{x}_o^{(\Delta m)} = \begin{pmatrix}
u_{o, m_-^{(\Delta m)}}^{(\Delta m)} \\ v_{o, m_-^{(\Delta m)}}^{(\Delta m)} \\ \vdots \\ u_{o, m_+^{(\Delta m)}}^{(\Delta m)} \\ v_{o, m_+^{(\Delta m)}}^{(\Delta m)}
\end{pmatrix}
\end{split},
\end{equation}
where $u_{o, m}^{(\Delta m)} \equiv \textrm{Re} (\rho_{(J_g,m),(J_e,m+\Delta m)} e^{- i \overline{\omega} t})$ and $v_{o, m}^{(\Delta m)} \equiv \textrm{Im} (\rho_{(J_g,m),(J_e,m+\Delta m)} e^{- i \overline{\omega} t})$, with $m = m_-^{(\Delta m)}, \ldots, m_+^{(\Delta m)}$, $m_\pm^{(\Delta m)} = \pm \min{(J_g, J_g+\Delta J \mp \Delta m)}$, and $\Delta m = -(J_e+J_g), \ldots, J_e+J_g$.

We defined $\mathbf{x}_p \equiv ( \mathbf{x}_{p_e}^T , \mathbf{x}_{p_g}^T )^T$, with ($k = e, g$)
\begin{equation}
\mathbf{x}_{p_k} = \begin{pmatrix} w_{k,-J_k+\delta_{k,g}} \\ \vdots \\ w_{k,J_k} \end{pmatrix} ,
\end{equation}
where $w_{k,m_k} = \rho_{(J_k, m_k),(J_k,m_k)} - N_J^{-1}$, with $m_k = -J_k + \delta_{k,g}, \ldots, J_k$ and $N_J \equiv 2(J_e+J_g+1)$.

We finally defined $\mathbf{x}_Z \equiv (\mathbf{x}_{Z_e}^T , \mathbf{x}_{Z_g}^T)^T$, with ($k = e, g$)
\begin{equation} \label{cohZee}
\begin{split}
\mathbf{x}_{Z_k} = \begin{pmatrix} \mathbf{x}_{Z_k}^{(1)} \\ \vdots \\ \mathbf{x}_{Z_k}^{(2J_k)} \end{pmatrix} , \textrm{ with } \mathbf{x}_{Z_k}^{(\Delta m)} = \begin{pmatrix}
u_{Z_k, -J_k}^{(\Delta m)} \\ v_{Z_k, -J_k}^{(\Delta m)} \\ \vdots \\ u_{Z_k, J_k-\Delta m}^{(\Delta m)} \\ v_{Z_k, J_k-\Delta m}^{(\Delta m)}
\end{pmatrix}
\end{split},
\end{equation}
where $u_{Z_k, m_k}^{(\Delta m)} = \textrm{Re} (\rho_{(J_k, m_k),(J_k, m_k + \Delta m)})$ and $v_{Z_k, m_k}^{(\Delta m)} = \textrm{Im} (\rho_{(J_k, m_k),(J_k, m_k+\Delta m)})$, with $m_k = -J_k, \ldots, J_k - \Delta m$ and $\Delta m = 1, \ldots , 2J_k$.

The generalized OBEs then immediately follow and read
\begin{equation} \label{MatrixOBE}
\mathbf{\dot{x}} (t) = A (t) \mathbf{x} (t) + \mathbf{b},
\end{equation}
where $A (t)$ and $\mathbf{b}$ are a matrix and a column vector as described in the next two subsections, respectively.

\subsubsection{The $A(t)$ matrix}

The $A(t)$ matrix reads
\begin{equation} \label{A0}
A (t) = - \Gamma A_0 + \textrm{Im} \left( \boldsymbol{\Omega} (t) \cdot \mathbf{e}_C \right),
\end{equation}
where $A_0$ is a time independent matrix with scalar entries (independent of any reference frame) detailed hereafter, $\boldsymbol{\Omega} (t) = \sum_j \boldsymbol{\Omega}_j(t) \equiv \sum_q \Omega_q (t) \mathbf{e}^q$, with $\boldsymbol{\Omega}_j(t) = \Omega_j e^{i \left( \omega_j - \overline{\omega} \right) t} \boldsymbol{\epsilon}_j$, where
\begin{equation}
\Omega_j = - \frac{E_j e^{i (- \mathbf{k}_j \cdot \mathbf{r} + \varphi_j)} }{\hbar} \frac{\langle J_e \| \mathbf{D} \| J_g \rangle^*}{\sqrt{2 J_e + 1}},
\end{equation}
in which $\langle J_e \| \mathbf{D} \| J_g \rangle$ denotes the so-called reduced matrix element associated to $\hat{\mathbf{D}}$ and $\mathbf{e}_C$ is the (unnormalized) basis vector of matrices $\mathbf{e}_C = \sum_q C^{(q)} \mathbf{e}_q$, with contravariant components $C^{(q)}$ as described hereafter and where $\mathbf{e}_q = {\mathbf{e}^q}^*$ ($q = 0, \pm 1$) are the lower-index spherical basis vectors ($\mathbf{e}_C$ is said a basis vector in that it rotates similarly with the spherical basis in case of a basis change, so that $\mathbf{\Omega}(t) \cdot \mathbf{e}_C = \sum_q \Omega_q(t) C^{(q)}$ does \emph{not} define a matrix of scalars - see Appendix~\ref{AppendixA}).

The $A_0$ matrix reads $A_0 = \textrm{diag}(A_{oo} , A_{\xi\xi})$, with matrix blocks $A_{oo} = \oplus^{(\dim{\mathbf{x}_o})/2} \Delta (1/2)$ and $A_{\xi\xi} = \textrm{diag} (A_{pp}, A_{ZZ})$, where
\begin{equation}
\Delta(z) = \begin{pmatrix} z & - \delta/\Gamma \\ \delta/\Gamma & z \end{pmatrix} , \quad \forall z \in \mathbb{C},
\end{equation}
and $A_{\zeta \zeta}$ ($\zeta = p, Z$) are blocks of dimension $\dim \mathbf{x}_\zeta \times \dim \mathbf{x}_\zeta$ themselves structured into subblocks according to
\begin{equation}
A_{\zeta\zeta} = \begin{pmatrix} \mathbb{1}_{\zeta_e} & 0 \\ A_{\zeta_g \zeta_e} & 0_{\zeta_g} \end{pmatrix},
\end{equation}
with, for any symbol $\alpha$, $\mathbb{1}_\alpha$ [$0_\alpha$] the identity [zero] matrix of dimension $\dim{\mathbf{x}_\alpha} \times \dim{\mathbf{x}_\alpha}$. The $A_{p_gp_e}$ subblock is of dimension $\dim \mathbf{x}_{p_g} \times \dim \mathbf{x}_{p_e}$ and of elements ${(A_{p_gp_e})_{m_gm_e} = - (\mathcal{C}_{m_g}^{(m_e-m_g)})^2}$, with $m_k = -J_k+\delta_{k,g},\ldots,J_k$ $(k = e, g)$ [here and throughout the paper, we adopt the convention not to index the matrix elements from $(1,1)$ but with indices directly linked to the magnetic sublevels]. The $A_{Z_gZ_e}$ subblock is of dimension $\dim \mathbf{x}_{Z_g} \times \dim \mathbf{x}_{Z_e}$ and is itself structured into vertically and  horizontally ordered subsubblocks $A_{Z_gZ_e}^{(\Delta m_g, \Delta m_e)}$ of dimension $\dim \mathbf{x}_{Z_g}^{(\Delta m_g)} \times \dim \mathbf{x}_{Z_e}^{(\Delta m_e)}$, with respective indices $\Delta m_g = 1, \ldots , 2J_g$ and $\Delta m_e = 1, \ldots, 2J_e$. The only \textit{a priori} nonzero of these subsubblocks are for $\Delta m_g = \Delta m_e \equiv \Delta m$, of elements $A_{Z_gZ_e}^{(\Delta m, \Delta m)} = \tilde{A}_{Z_gZ_e}^{(\Delta m, \Delta m)} \otimes \mathbb{1}_2$, with $(\tilde{A}_{Z_gZ_e}^{(\Delta m, \Delta m)})_{m_g m_e} = - \mathcal{C}_{m_g}^{(m_e - m_g)} \mathcal{C}_{m_g + \Delta m}^{(m_e - m_g)}$, where $m_k = -J_k , \ldots , J_k - \Delta m$ $(k = e, g)$.

The $C^{(q)}$ matrices read
\begin{equation} \label{CqMatrix}
C^{(q)} = \begin{pmatrix} 0 & C_{o \xi}^{(q)} \otimes (1, -i)^T \\ C_{\xi o}^{(q)} \otimes (1, -i) & 0 \end{pmatrix},
\end{equation}
with $C_{o \xi}^{(q)} = ( C_{op}^{(q)} , C_{oZ}^{(q)} )$ and $C_{\xi o}^{(q)} = ( {C_{po}^{(q)}}^T , {C_{Zo}^{(q)}}^T )^T$, where $C_{op}^{(q)} = ( C_{op_e}^{(q)} , C_{op_g}^{(q)} )$, $C_{po}^{(q)}= ( {C_{p_eo}^{(q)}}^T , {C_{p_go}^{(q)}}^T )^T$, $C_{oZ}^{(q)} = ( C_{oZ_e}^{(q)} , C_{oZ_g}^{(q)} )$, and $C_{Zo}^{(q)} = ( {C_{Z_eo}^{(q)}}^T , {C_{Z_go}^{(q)}}^T )^T$. The $C_{op_k}^{(q)}$, $C_{p_ko}^{(q)}$, $C_{oZ_k}^{(q)}$, and $C_{Z_ko}^{(q)}$ ($k = e, g$) blocks exclusively contain Clebsch-Gordan coefficients and are detailed below. In addition, for $\zeta = \xi, p_e, p_g, p, Z_e, Z_g, Z$, the $C_{o\zeta}^{(q)}$ [$C_{\zeta o}^{(q)}$] blocks are of dimension $(\dim{\mathbf{x}_o}/2) \times \dim{\mathbf{x}_\zeta}$ [$\dim{\mathbf{x}_\zeta} \times (\dim{\mathbf{x}_o}/2)$].

In accordance with Eq.~(\ref{cohOpt}), the $C_{op_k}^{(q)}$ blocks are structured into vertically ordered subblocks indexed with $\Delta m = -(J_g+J_e) , \ldots , J_g+J_e$ and of dimension $( \dim{\mathbf{x}_o^{(\Delta m)}}/2 ) \times \dim{\mathbf{x}_{p_k}}$. The only \textit{a priori} nonzero of these subblocks is for $\Delta m = q$ and we denote it by $\tilde{C}_{op_k}^{(q)}$. Its elements read $( \tilde{C}_{op_k}^{(q)} )_{m, m_k} = \mathcal{C}_{m}^{(q)} ( \delta_{m, -J_g} + \tilde{n}_k \delta_{m_k, m + n_k q} )/2$, with $n_k = \delta_{e,k}$ and $\tilde{n}_k = 2n_k-1$, where $m = m_-^{(q)} , \ldots , m_+^{(q)}$ and $m_k = -J_k + \delta_{k,g} , \ldots , J_k$. In a same way, the $C_{p_ko}^{(q)}$ blocks are structured into horizontally ordered subblocks indexed with $\Delta m = -(J_g+J_e) , \ldots , J_g+J_e$ and of dimension $\dim{\mathbf{x}_{p_k}} \times ( \dim{\mathbf{x}_o^{(\Delta m)}}/2 )$. Again, the only \textit{a priori} nonzero of these subblocks is for $\Delta m = q$ and it is denoted by $\tilde{C}_{p_ko}^{(q)}$. Its elements read $( \tilde{C}_{p_ko}^{(q)} )_{m_k, m} = - \tilde{n}_k \mathcal{C}_{m}^{(q)} \delta_{m, m_k - n_k q}$, where $m_k = -J_k + \delta_{k,g} , \ldots , J_k$ and $m = m_-^{(q)} , \ldots , m_+^{(q)}$.

Finally, $C_{Z_ko}^{(q)} = - {C_{oZ_k}^{(q)}}^T$ and $C_{oZ_k}^{(q)} = \sum_{\epsilon = \pm 1} C_{oZ_k, \epsilon}^{(q)} \otimes (1, \epsilon i)$, where, in accordance with Eqs.~(\ref{cohOpt}) and~(\ref{cohZee}), $C_{oZ_k, \epsilon}^{(q)}$ is structured into vertically and horizontally ordered subblocks, with respective indices $\Delta m = -(J_g+J_e), \ldots, J_g+J_e$ and $\Delta m_k = 1, \ldots, 2J_k$. These subblocks are of dimension $( \dim{\mathbf{x}_o^{(\Delta m)}}/2 ) \times ( \dim{\mathbf{x}_{Z_k}^{(\Delta m_k)}}/2 )$. The only \textit{a priori} nonzero of them are for $\Delta m - \epsilon \Delta m_k = q$ and they are denoted by $\tilde{C}_{oZ_k, \epsilon}^{(q)(\Delta m_k)}$. Their elements read $( \tilde{C}_{oZ_k, \epsilon}^{(q)(\Delta m_k)} )_{m, m_k} = (\tilde{n}_k/2) \mathcal{C}_{m - \epsilon (n_k - 1) \Delta m_k}^{(q)} \delta_{m_k, m + n_k q - (1-\epsilon)\Delta m_k/2}$, where $m = m_-^{(\Delta m)} , \ldots , m_+^{(\Delta m)}$ and $m_k = -J_k, \ldots, J_k - \Delta m_k$.

\subsubsection{The $\mathbf{b}$ column vector}

With the convention of using a $\zeta$ index to denote a $\dim{\mathbf{x}_\zeta}$ column vector, the $\mathbf{b}$ column vector reads $\mathbf{b} = -\Gamma N_J^{-1} ( \mathbf{b}_o^T , \mathbf{b}_\xi^T  )^T$, with $\mathbf{b}_o = 0$ and $\mathbf{b}_\xi = A_{\xi \xi} \mathbf{u}_\xi$. Here, $\mathbf{u}_\xi = (\mathbf{u}_p^T,\mathbf{u}_Z^T)^T$, with $\mathbf{u}_Z = 0$ and $\mathbf{u}_p = (\mathbf{u}_{p_e}^T,\mathbf{u}_{p_g}^T)^T$, where $\mathbf{u}_{p_g} = 0$ and $\mathbf{u}_{p_e}$ is a column vector only composed of $1$.

\subsection{Periodic regime}

The OBEs (\ref{MatrixOBE}) cannot be solved analytically and require numerical integration in the most general case. However, within the commensurability assumption (all $\omega_j - \overline{\omega}$ commensurable), the $\boldsymbol{\Omega} (t)$ and $A(t)$ quantities are periodic in time with the repetition period $T_c = 2 \pi / \omega_c$, where $\omega_c = ( \textrm{LCM} [ (\omega_j - \overline{\omega})^{-1} , \forall j : \omega_j \ne \overline{\omega} ] )^{-1}$, and all $m_j = (\omega_j - \overline{\omega})/\omega_c$ numbers are integer numbers. In this case, the OBEs admit in all circumstances a $T_c$-periodic solution since $\mathbf{x}(t)$ is bounded~\cite{Adrianova1995}. This solution is unique if the real parts $\lambda$ of the Floquet exponents are all strictly negative, in which case the system necessarily converges to it (and reaches a so-called periodic regime) from any initial state $\mathbf{x}(t_0)$ after a transient of characteristic time $\vert \lambda_m \vert^{-1}$, with $\lambda_m$ the greatest $\lambda$~\cite{Adrianova1995}. For two-level atoms, $-\Gamma \leq \lambda \leq -\Gamma/2$ and the transient characteristic time remains limited to at most $2\Gamma^{-1}$~\cite{Podlecki2018}. This is not the case anymore for multilevel atoms where greater damping times can be observed. It might even happen that $\lambda = 0$, in which case several $T_c$-periodic solutions might exist (the Floquet exponents must be computed numerically and we only observed $\lambda = 0$ in specific configurations with $\Delta J = -1$). The case $\lambda > 0$ can never happen since this would correspond to unphysical unbounded $\mathbf{x}(t)$.

Any $T_c$-periodic solution $\mathbf{x}(t)$ can be expanded according to
\begin{equation} \label{FourierDecompx}
\mathbf{x} (t) = \sum\limits_{n = - \infty}^{+ \infty} \mathbf{x}^{(n)} e^{i n \omega_c t},
\end{equation}
with Fourier components $\mathbf{x}^{(n)}$. The variable $\mathbf{x} (t)$ is real, continuous, and differentiable, so that $\mathbf{x}^{(-n)} = {\mathbf{x}^{(n)}}^*$ and $\sum_n \vert \mathbf{x}^{(n)} \vert^2 < \infty$. For $\zeta = o, \xi, p, Z, p_e, p_g, Z_e, Z_g$, we denote by $\mathbf{x}_\zeta^{(n)}$ the Fourier components related to the periodic variable $\mathbf{x}_\zeta (t)$. Inserting Eq.~(\ref{FourierDecompx}) into (\ref{MatrixOBE}) yields the infinite system of equations
\begin{equation} \label{SolFouriero}
\mathbf{x}_o^{(n)} = A_{oo}^{(n)} \sum\limits_{q, j}  \left( \check{C}_{o \xi}^{(q)^{\scriptstyle{*}}} \frac{\Omega_{j, q}^*}{\Gamma} \mathbf{x}_\xi^{(n+m_j)} - \check{C}_{o \xi}^{(q)}  \frac{\Omega_{j, q}}{\Gamma} \mathbf{x}_\xi^{(n - m_j)} \right), \forall n,
\end{equation}
with $\Omega_{j,q} = \Omega_j \epsilon_{j,q}$, $A_{oo}^{(n)} = - 2i \Gamma^2 \tau_n^+ \tau_n^- \mathbb{1}_{\dim{\mathbf{x}_o}/2} \otimes \Delta (- i n \omega_c/\Gamma - 1/2)$, and $\check{C}_{o \xi}^{(q)} = C_{o \xi}^{(q)} \otimes (1, -i)^T$ [see Eq.~(\ref{CqMatrix})], where we defined $\tau_n^\pm = 1/[\Gamma + 2 i (n \omega_c \pm \overline{\delta})]$, and
\begin{equation} \label{FinalSystFourier}
\mathbf{x}_\xi^{(n)} + \sum\limits_{m \in M_0} \mathcal{W}_{\xi \xi}^{(n, m)} \mathbf{x}_\xi^{(n+m)} = \mathbf{d}_\xi \delta_{n, 0}, \forall n.
\end{equation}
In the latter, $\delta_{n,0}$ denotes the Kronecker symbol, $M_0$ is the set of all distinct nonzero integer numbers $m_{lj} \equiv m_l-m_j$ $(j, l = 1, \ldots, N)$, and we defined $\mathcal{W}_{\xi \xi}^{(n, m)} = ( A_{\xi\xi}^{(n)} + B_{\xi\xi}^{(n, 0)} )^{-1} B_{\xi \xi}^{(n, m)}$ and $\mathbf{d}_\xi = -N_J^{-1} ( A_{\xi\xi} + \tilde{s}_{\xi \xi} )^{-1} A_{\xi \xi} \mathbf{u}_\xi$, where we set $A_{\xi\xi}^{(n)} = A_{\xi\xi} + (i n \omega_c/\Gamma) \mathbb{1}_\xi$, $B_{\xi\xi}^{(n, m)} = \sum_{j, l : m_{lj} = m} \sum_{q,q^\prime} (\Omega_{j, q} \Omega_{l, q^\prime}^*/\Gamma) ( \tau_{n - m_j}^- (C_{\xi\xi})^q_{q^\prime} + \tau_{n+m_l}^+ {(C_{\xi\xi})_q^{q^{\prime}}}^* )$ with $(C_{\xi\xi})_{q^\prime}^q \equiv - C_{\xi o}^{(q)} {C_{o \xi}^{(q^\prime)}}^*$, and $\tilde{s}_{\xi \xi} = \sum_j \tilde{s}_{\xi \xi, j}$ with
\begin{equation}
\tilde{s}_{\xi\xi,j} = \textrm{Re} \left[ \sum\limits_{q,q^\prime} \sum\limits_{\underset{l : m_{lj} = 0}{l=1}}^N  \frac{\Omega_{j, q} \Omega_{l, q^\prime}^*/\Gamma}{\Gamma/2 - i \delta_j} (C_{\xi\xi})^q_{q^\prime} \right] .
\end{equation}
We have $\tau_{-n}^\pm = {\tau_n^{\mp}}^*$, $A_{\xi\xi}^{(-n)} = {A_{\xi\xi}^{(n)}}^*$, $B_{\xi\xi}^{(-n,-m)} = {B_{\xi\xi}^{(n,m)}}^*$, and $\mathcal{W}_{\xi\xi}^{(-n,-m)} = {\mathcal{W}_{\xi\xi}^{(n,m)}}^*$.

The system~(\ref{FinalSystFourier}) is only defined if all $A_{\xi \xi}^{(n)} + B_{\xi \xi}^{(n, 0)}$ matrices are invertible, which is ensured if the periodic regime is unique. Otherwise, the following formalism does not apply. This is for instance the case for $\Delta J = -1$ and all waves with the same polarization $\boldsymbol{\epsilon}_j$ ($\lambda_m = 0$ and $\det{(A_{\xi \xi}^{(0)} + B_{\xi \xi}^{(0, 0)})} = 0$).

If we define $\mathbf{y} = ( \ldots , {\mathbf{x}_\xi^{(-1)}}^T , {\mathbf{x}_\xi^{(0)}}^T , {\mathbf{x}_\xi^{(1)}}^T , \ldots )^T$ the infinite column vector of all $\mathbf{x}_\xi^{(n)}$ components, as well as
\begin{equation} \label{DefInfiniteWMulti}
W = \begin{pmatrix}
\ddots & \vdots & \vdots & \vdots & \udots \\
\ldots & W_{\xi \xi}^{(-1,-1)} & W_{\xi \xi}^{(-1,0)} & W_{\xi \xi}^{(-1,1)} & \ldots \\
\ldots & W_{\xi \xi}^{(0,-1)} & W_{\xi \xi}^{(0,0)} & W_{\xi \xi}^{(0,1)} & \ldots \\
\ldots & W_{\xi \xi}^{(1,-1)} & W_{\xi \xi}^{(1,0)} & W_{\xi \xi}^{(1,1)} & \ldots \\
\udots & \vdots & \vdots & \vdots & \ddots \\
\end{pmatrix}
\end{equation}
the infinite matrix structured into the subblocks $W_{\xi \xi}^{(n, n^\prime)} = \sum_{m \in M_0} \mathcal{W}_{\xi \xi}^{(n,m)} \delta_{n^\prime, n+m}$ ($n$, $n^\prime$ ranging from $- \infty$ to $+ \infty$), then Eq.~(\ref{FinalSystFourier}) yields the complex inhomogeneous infinite system of equations
\begin{equation} \label{InfSystemWyc}
(I + W) \mathbf{y} = \mathbf{c},
\end{equation}
with $\mathbf{c} = ( \ldots , {\mathbf{c}_\xi^{(-1)}}^T , {\mathbf{c}_\xi^{(0)}}^T , {\mathbf{c}_\xi^{(1)}}^T , \ldots )^T$ the infinite column vector of subblocks $\mathbf{c}_\xi^{(n)} = \mathbf{d}_\xi \delta_{n,0}$ ($n$ ranging from $- \infty$ to $+ \infty$) and $I$ the infinite identity matrix. We directly have $W_{\xi \xi}^{(n, n)} = 0$, $\forall n$. The solution of the system~(\ref{InfSystemWyc}) is necessarily such as $\mathbf{x}_\xi^{(0)} \neq 0$; otherwise, all other $\mathbf{x}_\xi^{(n)}$ components would solve a homogeneous system of equations and thus vanish, in which case the equation for $n=0$ could not be satisfied. This allows us to define the matrices $Q_{\xi \xi}^{(n)}$ that map the $\mathbf{x}_\xi^{(0)}$ vector onto $\mathbf{x}_\xi^{(n)}$, $\forall n$: $Q_{\xi \xi}^{(n)} \mathbf{x}_\xi^{(0)} = \mathbf{x}_\xi^{(n)}$. The $Q_{\xi \xi}^{(n)}$ matrices are \textit{a priori} not unique. Such suitable matrices can be given by (see, e.g., Ref.~\cite{Horn2012}): $Q_{\xi \xi}^{(n)} = 0$ if $\mathbf{x}_\xi^{(n)} = 0$; otherwise $Q_{\xi \xi}^{(n)} = (e^{i \theta} \| \mathbf{x}_\xi^{(n)} \| /\| \mathbf{x}_\xi^{(0)} \|) \mathbb{1}_\xi$ if $\mathbf{x}_\xi^{(n)} / \| \mathbf{x}_\xi^{(n)} \| = e^{i \theta} (\mathbf{x}_\xi^{(0)} / \| \mathbf{x}_\xi^{(0)} \|)$ ($\theta \in [ 0, 2 \pi [$); otherwise $Q_{\xi \xi}^{(n)} = (\| \mathbf{x}_\xi^{(n)} \|/\| \mathbf{x}_\xi^{(0)} \|) U_{\xi \xi}^{(n)}$. In the latter, $U_{\xi \xi}^{(n)}$ is the unitary matrix $e^{i \phi^{(n)}} V_{\xi \xi}^{(n)}$, where $\phi^{(n)}$ is the phase of the complex number ${\mathbf{x}_\xi^{(0)}}^* \cdot \mathbf{x}_\xi^{(n)}$, conventionally set to 0 if the complex number is zero, and where $V_{\xi \xi}^{(n)}$ is the Householder matrix $V_{\xi \xi}^{(n)} = \mathbb{1}_\xi - (2/\| \mathbf{z}_\xi^{(n)} \|^2) ( \mathbf{z}_\xi^{(n)} {\mathbf{z}_\xi^{(n)}}^\dagger )$, with $\mathbf{z}_\xi^{(n)} = e^{i \phi^{(n)}} \mathbf{x}_\xi^{(0)} - (\| \mathbf{x}_\xi^{(0)} \|/\| \mathbf{x}_\xi^{(n)} \|) \mathbf{x}_\xi^{(n)}$. In particular, we have in that case $Q_{\xi\xi}^{(0)} = \mathbb{1}_\xi$ and $Q_{\xi\xi}^{(-n)} = {Q_{\xi\xi}^{(n)}}^*$. Inserting $\mathbf{x}_\xi^{(n)} = Q_{\xi \xi}^{(n)} \mathbf{x}_\xi^{(0)}$ into Eq.~(\ref{FinalSystFourier}) for $n=0$ yields
\begin{equation} \label{xxi0Last}
\mathbf{x}_\xi^{(0)} = - \frac{1}{N_J} \frac{A_{\xi \xi}}{A_{\xi \xi} + s_{\xi \xi}} \mathbf{u}_\xi,
\end{equation}
where $\frac{A_{\xi \xi}}{A_{\xi \xi} + s_{\xi \xi}} \equiv \left( A_{\xi \xi} + s_{\xi \xi} \right)^{-1} A_{\xi \xi}$ and $s_{\xi \xi} = \sum_j s_{\xi \xi, j}$, with
\begin{equation} \label{SatParam}
s_{\xi \xi,j} = \textrm{Re} \left[ \sum\limits_{q,q^\prime} \sum\limits_{l=1}^N \frac{\Omega_{j, q} \Omega_{l, q^\prime}^*/\Gamma}{\Gamma/2 - i \delta_j} (C_{\xi\xi})^q_{q^\prime} Q_{\xi \xi}^{(m_{lj})} \right] .
\end{equation}
Equation~(\ref{SatParam}) is the generalization of Eq.~(15) of Ref.~\cite{Podlecki2018} to the degenerate two-level atom case.

\subsection{General and exact expression of the radiation force}

Proceeding along the same lines as in Ref.~\cite{Podlecki2018}, the total mean power absorbed from all plane waves and the mean net force exerted on the atom can be expressed as $P(t) = \sum_j P_j(t)$ and $\mathbf{F}(t) = \sum_j \mathbf{F}_j(t)$, respectively, with $P_j (t) = R_j (t) \hbar \overline{\omega}$ and $\mathbf{F}_j (t) = R_j (t) \hbar \mathbf{k}_j$, where
\begin{equation}
R_j (t) = \textrm{Im} \left[\boldsymbol{\Omega}_j(t) \cdot \boldsymbol{\chi}_o(t)\right],
\label{chijt}
\end{equation}
with $\boldsymbol{\chi}_o(t) \equiv \sum_q \chi^{(q)}_o(t) \mathbf{e}_q$ the three-dimensional vector of contravariant components
\begin{equation}
\chi^{(q)}_o(t) = - \sum_{m = m_-^{(q)}}^{m_+^{(q)}} \mathcal{C}_{m}^{(q)}(u_{o, m}^{(q)}(t) - i v_{o, m}^{(q)}(t)).
\label{chiqt2}
\end{equation}
The vector $\boldsymbol{\chi}_o(t)$ is a true polar vector (see Appendix~\ref{AppendixA}), so that obviously $\boldsymbol{\Omega}_j(t) \cdot \boldsymbol{\chi}_o(t)$ is a scalar quantity. As a reminder~\cite{Podlecki2018}, $R_j(t)$ merely yields in the quasi-resonance condition the mean photon absorption rate, $\langle dN/dt \rangle_j(t)$, induced by plane wave $j$.

In a periodic regime, $R_{j}(t)$ can be expanded according to $\sum_{n = - \infty}^{+\infty} R_j^{(n)} e^{i n \omega_c t}$, with Fourier components $R_j^{(n)}$. When this regime is unique, these components are obtained by inserting Eq.~(\ref{SolFouriero}) into Eq.~(\ref{chijt}) and considering $Q_{\xi \xi}^{(n)} \mathbf{x}_\xi^{(0)} = \mathbf{x}_\xi^{(n)}$, along  with $\mathbf{x}_\xi^{(0)}$ as given by Eq.~(\ref{xxi0Last}). This yields
\begin{equation} \label{ExpressionRjn}
R_j^{(n)} = \frac{\Gamma}{N_J} \mathbf{u}_\xi^T s_{\xi \xi, j}^{(n)} \frac{A_{\xi \xi}}{A_{\xi \xi} + s_{\xi \xi}} \mathbf{u}_\xi = - \Gamma \mathbf{u}_\xi^T s_{\xi\xi,j}^{(n)} \mathbf{x}_\xi^{(0)},
\end{equation}
where $s_{\xi \xi, j}^{(n)} = \left( \sigma_{\xi\xi, j}^{(n)} + {\sigma_{\xi\xi, j}^{(-n)}}^* \right)/2$, with
\begin{equation} \label{Defsigmaxixijn}
\sigma_{\xi \xi, j}^{(n)} = \sum\limits_{q, q^\prime} \sum\limits_{l=1}^N \frac{\Omega_{j, q} \Omega_{l, q^\prime}^*/\Gamma}{\Gamma/2 + i (n \omega_c - \delta_j)} (C_{\xi \xi})^q_{q^\prime}  Q_{\xi \xi}^{(n+ m_{lj})} .
\end{equation}
In particular, observing that $s_{\xi\xi,j}^{(0)} = s_{\xi\xi,j}$ [Eq.~(\ref{SatParam})], the temporal mean value $\overline{R}_j \equiv R_j^{(0)}$ of $R_{j}(t)$ in the unique periodic regime reads
\begin{equation} \label{Rj0}
\overline{R}_j = \frac{\Gamma}{N_J} \mathbf{u}_\xi^T s_{\xi \xi, j} \frac{A_{\xi \xi}}{A_{\xi \xi} + s_{\xi \xi}} \mathbf{u}_\xi
\end{equation}
and the corresponding mean force similarly reads
\begin{equation} \label{finalFj}
\overline{\mathbf{F}}_j = \frac{\Gamma}{N_J} (\mathbf{u}_\xi^T s_{\xi \xi, j} \frac{A_{\xi \xi}}{A_{\xi \xi} + s_{\xi \xi}} \mathbf{u}_\xi)  \hbar \mathbf{k}_j .
\end{equation}
Equation~(\ref{finalFj}) is a natural extension of Eq.~(19) of Ref.~\cite{Podlecki2018}. Interestingly, the two-level atom case is also covered within the present formalism. To this aim, it is enough to consider $J_g = J_e = 0$ and to open artificially the forbidden 0-0 transition by forcing $\mathcal{C}_0^{(0)}$ to 1. All equations above then merely simplify to the two-level atom formalism of Ref.~\cite{Podlecki2018}.

\begin{figure}
\begin{centering}
\includegraphics[width=0.4\textwidth]{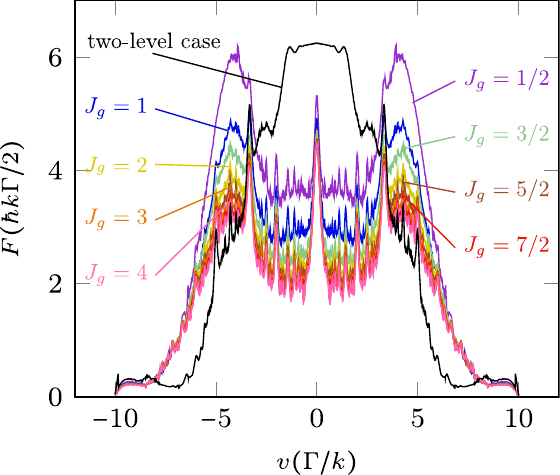}
\caption{(Color online) Stimulated bichromatic force $F$ as a function of the atomic velocity $v$ computed via our formalism for a detuning $\delta = 10 \Gamma$, a Rabi frequency of $\sqrt{3/2}\delta$, a phase shift of $\pi/2$ for one wave, $\boldsymbol{\epsilon}_j = \boldsymbol{\pi}, \forall j$, and $J_g = 1/2 , 1 , \ldots , 4$ with $\Delta J = 1$.}
\label{FigBichr}
\end{centering}
\end{figure}

To illustrate our formalism, we show how the stimulated bichromatic force in a standard four traveling-wave configuration~\cite{Soding1997} compares between the two-level~\cite{Podlecki2018} and multilevel atom cases for several atomic structures. Light with pure $\pi$ polarization ($\boldsymbol{\epsilon}_j = \boldsymbol{\pi}, \forall j$) was considered for the comparison. We show in Fig.~\ref{FigBichr} the value of the resulting bichromatic force (averaged over the $2 \pi$ range of the spatially varying relative phase between the opposite waves) acting in the direction of the phased wave on a moving atom as a function of its velocity $v$ for $\Delta J = 1$ with $J_g = 1/2, 1, \ldots, 4$. Figure~\ref{FigBichr} shows that the $\pi$ polarization case yields poorer results than the ideal two-level atom case.

\section{Specific regimes} \label{SecSpecific}

The general expression of the mean photon absorption rate (\ref{Rj0}) can be interestingly simplified in several specific regimes. We investigate some emblematic cases in the next subsections.

\subsection{Low-intensity regime}

We define the low-intensity regime as the regime where $\sum_j \vert \Omega_j \vert / \Gamma \ll 1$ and~\cite{LowIntRegime}
\begin{equation} \label{LowIntensityCond}
\sum\limits_{m_2} \sum\limits_{m \in M_0} \vert ( \mathcal{W}_{\xi \xi}^{(n, m)} )_{m_1, m_2} \vert \ll \frac{1}{\sqrt{\dim{\mathbf{x}_\xi}}} , \quad \forall n \neq 0 , \forall m_1 .
\end{equation}
In this regime, we have $ s_{\xi \xi, j} \simeq \tilde{s}_{\xi \xi, j}$ and $\overline{R}_j$ can be computed without solving the infinite system~(\ref{InfSystemWyc}). Indeed, Eq.~(\ref{InfSystemWyc}) can be expressed as a function of $\mathbf{x}_\xi^{(0)}$ according to $( \mathbb{1} + W_0 ) \mathbf{y}_0 = - W_0^\prime \mathbf{x}_\xi^{(0)}$, where $\mathbf{y}_0$ is the $\mathbf{y}$ vector excluding the $\mathbf{x}_\xi^{(0)}$ component, $W_0$ is the $W$ matrix excluding the $W_{\xi \xi}^{(0, n^\prime)}$ and $W_{\xi \xi}^{(n, 0)}$ blocks, $\forall n , n^\prime$, and $W_0^\prime$ is the $W$ matrix restricted to the only blocks $W_{\xi \xi}^{(n, 0)}$, $\forall n \neq 0$. Under condition~(\ref{LowIntensityCond}), we have $(\mathbb{1} + W_0)^{-1} \simeq \mathbb{1} + \sum_{k=1}^{+ \infty} (-W_0)^k$. This implies $\mathbf{y}_0 \simeq - W_0^\prime \mathbf{x}_\xi^{(0)}$, so that $\| \mathbf{x}_\xi^{(n)} \| \ll \| \mathbf{x}_\xi^{(0)} \|$ and thus $Q_{\xi \xi}^{(n)} \simeq 0_{\xi \xi} , \forall n \neq 0$, i.e., $ s_{\xi \xi, j} \simeq \tilde{s}_{\xi \xi, j}$. With all distinct frequencies, we get
\begin{equation} \label{sxixijLowAndDiffFreqMulti}
s_{\xi \xi, j} \simeq \textrm{Re} \left[ \sum\limits_{q,q^\prime} \left( \frac{\Omega_{j, q} \Omega_{j, q^\prime}^* / \Gamma}{\Gamma/2 - i \delta_j} \right) (C_{\xi\xi})^q_{q^\prime} \right].
\end{equation}
Otherwise coherent effects can be observed. The only incoherent contribution is obtained with an averaging over all phase differences. For $N=2$ with $\boldsymbol{\epsilon}_1 = \boldsymbol{\sigma}^+$ and $\boldsymbol{\epsilon}_2 = \boldsymbol{\sigma}^-$, we get in particular $\overline{R}_j^{\textrm{inc}} \equiv \langle \overline{R}_j \rangle_\varphi \simeq \Gamma N_J^{-1} \mathbf{u}_\xi^T \langle \tilde{s}_{\xi\xi, j} \rangle_\varphi \left( A_{\xi\xi} + \sum_i \langle \tilde{s}_{\xi\xi, i} \rangle_\varphi \right)^{-1} A_{\xi\xi} \mathbf{u}_\xi$. For such polarization vectors, we further have $\langle \tilde{s}_{\xi\xi, j} \rangle_\varphi = \textrm{Re} [ (\vert \Omega_j\vert^2/\Gamma) / (\Gamma/2 - i \delta_j) (C_{\xi\xi})^{q_j}_{q_j} ]$, where $q_1 = 1 = -q_2$. Since $(C_{\xi\xi})^q_q$ has the block-diagonal structure (see Appendix~\ref{AppendixB})
\begin{equation} \label{BlocdiagCxixiqq}
(C_{\xi\xi})^q_q = \begin{pmatrix} (C_{pp})^q_q & 0 \\ 0 & (C_{ZZ})^q_q \end{pmatrix} ,
\end{equation}
with $(C_{\zeta \zeta})^q_q = - C_{\zeta o}^{(q)} C_{o \zeta}^{(q)^{\scriptstyle{*}}}$ ($\zeta=p,Z$), it follows that $\langle \tilde{s}_{\xi\xi, j} \rangle_\varphi$ has the same block-diagonal structure. In addition, $A_{\xi\xi}$ behaves similarly and since all $(C_{pp})^q_q$ matrix elements are real numbers, $\overline{R}_j^{\textrm{inc}}$ simplifies to $\overline{R}_j^{\textrm{inc}} \simeq \Gamma N_J^{-1} s_j f_j(s_1,s_2)$, with $s_j \equiv (\vert \Omega_j \vert^2/2)/(\Gamma^2/4 + \delta_j^2)$ and $f_j(s_1,s_2) = \mathbf{u}_p^T (C_{pp})^{q_j}_{q_j} ( A_{pp} + \sum_{i=1}^2 s_i (C_{pp})^{q_i}_{q_i} )^{-1} A_{pp} \mathbf{u}_p$.

\subsection{Plane waves with same frequency}

In this case, any periodic regime is a strict stationary regime ($R_j^{(n)} = 0, \forall n \neq 0$) and $s_{\xi \xi, j}$ [see Eq.~(\ref{SatParam})] simplifies to $\textrm{Re} [ \sum_{q, q^\prime} (\underline{s}_j)_q^{q^\prime} (C_{\xi\xi})^q_{q^\prime} ]$, where $(\underline{s}_j)_q^{q^\prime}$ is the second-order tensor $(\Omega_{j, q} \Omega_{q^\prime}^*/\Gamma)/(\Gamma/2 - i \delta)$, with $\delta \equiv \delta_j, \forall j$. Here, $\Omega_q$ is independent of time and merely identifies to $\sum_j \Omega_{j,q}$.

For $N=1$, the index $j = 1$ can be omitted and we get $\underline{s}_q^{q^\prime} = s (1+2i\delta/\Gamma) \epsilon_q \epsilon_{q^\prime}^*$ with $s = (\vert \Omega \vert^2/2)/(\Gamma^2/4+\delta^2)$ and $\overline{R} = \Gamma N_J^{-1} \mathbf{u}_\xi^T f_{\xi\xi} ( A_{\xi \xi} + f_{\xi\xi} )^{-1} A_{\xi \xi} \mathbf{u}_\xi$ where we set $f_{\xi\xi} \equiv \textrm{Re} [ \sum_{q, q^\prime} \underline{s}_q^{q^\prime} (C_{\xi\xi})^q_{q^\prime} ]$. For a given atomic structure and in contrast to $A_{\xi\xi}$ and $\mathbf{u}_\xi$, the $f_{\xi\xi}$ matrix and $\overline{R}$ depend on $s$, $\boldsymbol{\epsilon}$, and $\delta$ (at constant $s$): $f_{\xi\xi} \equiv f_{\xi\xi} (s, \boldsymbol{\epsilon}, \delta)$ and $\overline{R} \equiv \overline{R} (s, \boldsymbol{\epsilon}, \delta)$. In addition, we have $f_{\xi\xi} (s, \boldsymbol{\epsilon}, \delta) = f_{\xi\xi} (s, \boldsymbol{\epsilon}, 0) - (2 \delta s / \Gamma) I_{\xi\xi} (\boldsymbol{\epsilon})$ with $I_{\xi\xi} (\boldsymbol{\epsilon}) = \textrm{Im} [ \sum_{q, q^\prime} \epsilon_q \epsilon_{q^\prime}^* (C_{\xi\xi})^q_{q^\prime} ]$. Since, for all invertible matrices $A$ and $A-B$, we have $(A-B)^{-1} = A^{-1} + (A-B)^{-1} B A^{-1}$, we get $X_{\xi\xi} (s, \boldsymbol{\epsilon}, \delta)^{-1} = X_{\xi\xi} (s, \boldsymbol{\epsilon}, 0)^{-1} + (2 \delta s/\Gamma) X_{\xi\xi} (s, \boldsymbol{\epsilon}, \delta)^{-1} I_{\xi\xi} (\boldsymbol{\epsilon}) X_{\xi\xi} (s, \boldsymbol{\epsilon}, 0)^{-1}$, where we set $X_{\xi\xi} (s, \boldsymbol{\epsilon}, \delta) \equiv A_{\xi\xi} + f_{\xi\xi} (s, \boldsymbol{\epsilon}, \delta)$. It follows that $\overline{R} (s, \boldsymbol{\epsilon}, \delta) = \overline{R} (s, \boldsymbol{\epsilon}, 0) - 2 \delta s N_J^{-1} \mathbf{u}_\xi^T A_{\xi\xi} X_{\xi\xi} (s, \boldsymbol{\epsilon}, \delta)^{-1} \mathbf{u}^\prime_{\xi} (s, \boldsymbol{\epsilon})$, where $\mathbf{u}^\prime_{\xi} (s, \boldsymbol{\epsilon}) \equiv I_{\xi\xi} (\boldsymbol{\epsilon}) X_{\xi\xi} (s, \boldsymbol{\epsilon}, 0)^{-1} A_{\xi\xi} \mathbf{u}_\xi$. The graph of $\| \mathbf{u}^\prime_\xi (s, \boldsymbol{\epsilon}) \|^2$ as a function of the most general polarization configuration $\boldsymbol{\epsilon} = \cos(\theta/2) \boldsymbol{\sigma}^+ + e^{i \varphi} \sin(\theta/2) \boldsymbol{\sigma}^-$ ($\theta \in [0,\pi], \varphi \in [0,2\pi[$) always identifies to 0, whatever $s$ and the atomic structure ($J_g, J_e$), so that
\begin{equation} \label{RIndepdelta}
\overline{R} (s, \boldsymbol{\epsilon}, \delta) = \overline{R} (s, \boldsymbol{\epsilon}, 0) ,
\end{equation}
i.e., $\overline{R}$ does not depend on $\delta$ at constant $s$.

\subsection{Plane waves with same pure polarization}

If all plane waves have the same pure polarization $q$ ($\boldsymbol{\epsilon}_j = \mathbf{e}^q$, $\forall j$), then each $\mathcal{W}_{\xi \xi}^{(n, m)}$ matrix is block-diagonal with blocks $\mathcal{W}_{pp}^{(n, m)}$ and $\mathcal{W}_{ZZ}^{(n, m)}$ of dimension $\dim{\mathbf{x}_p} \times \dim{\mathbf{x}_p}$ and $\dim{\mathbf{x}_Z} \times \dim{\mathbf{x}_Z}$, respectively, because so are the $(C_{\xi\xi})^q_q$ matrices, with corresponding blocks $(C_{pp})^q_q$ and $(C_{ZZ})^q_q$. In addition, the $\dim{\mathbf{x}_Z}$ last components of the $\mathbf{d}_\xi$ vector vanish and its $\dim{\mathbf{x}_p}$ first components are denoted hereafter by $\mathbf{d}_p$. Equation~(\ref{FinalSystFourier}) then yields the decoupled system
\begin{equation}
\left\{ \begin{split}
\mathbf{x}_p^{(n)} + \sum\limits_{m \in M_0} \mathcal{W}_{pp}^{(n, m)} \mathbf{x}_p^{(n+m)} & = \mathbf{d}_p \delta_{n, 0} , \\
\mathbf{x}_Z^{(n)} + \sum\limits_{m \in M_0} \mathcal{W}_{ZZ}^{(n, m)} \mathbf{x}_Z^{(n+m)} & = 0
\end{split} \right.
\end{equation}
that can be solved separately for the $\mathbf{x}_p^{(n)}$ and $\mathbf{x}_Z^{(n)}$ Fourier components. The $\mathbf{x}_Z^{(n)}$ components satisfy a homogeneous system with trivial solution $\mathbf{x}_Z^{(n)} = 0, \forall n$. This implies that the $\dim{\mathbf{x}_Z}$ last components of $\mathbf{z}_\xi^{(n)}$ are zero and then the $Q_{\xi\xi}^{(n)}$ matrix is block-diagonal with blocks $Q_{pp}^{(n)}$ and $Q_{ZZ}^{(n)}$ of dimension $\dim{\mathbf{x}_p} \times \dim{\mathbf{x}_p}$ and $\dim{\mathbf{x}_Z} \times \dim{\mathbf{x}_Z}$, respectively. As a consequence, $\sigma_{\xi \xi, j}^{(n)}$ are in turn block-diagonal with the same structure, $\forall n$, and so are $s_{\xi\xi,j}^{(n)}$ with corresponding blocks $s_{pp,j}^{(n)}$ and $s_{ZZ,j}^{(n)}$. In particular, Eqs.~(\ref{xxi0Last}) and~(\ref{ExpressionRjn}) yield $\mathbf{x}_p^{(0)} = - N_J^{-1} (A_{pp} + s_{pp})^{-1} A_{pp} \mathbf{u}_p$ and $R_j^{(n)} = \Gamma N_J^{-1} \mathbf{u}_p^T s_{pp, j}^{(n)} \left( A_{pp} + s_{pp} \right)^{-1} A_{pp} \mathbf{u}_p$, respectively, with $s_{pp} = \sum_j s_{pp, j}$, where $s_{pp, j} \equiv s_{pp, j}^{(0)}$.

In the low-intensity regime and with all distinct frequencies, we get $s_{pp, j} \simeq s_j (C_{pp})_q^q$ and $\overline{R}_j = \Gamma N_J^{-1} s_j \mathbf{u}_p^T (C_{pp})_q^q (A_{pp} + s (C_{pp})_q^q)^{-1} A_{pp} \mathbf{u}_p$, respectively, with $s = \sum_j s_j$ and $s_j = (\vert \Omega_j \vert^2/2)/(\Gamma^2/4 + \delta_j^2)$.

\subsection{Plane waves with same frequency and pure polarization}

\begin{figure}
\begin{centering}
\includegraphics[width=0.48\textwidth]{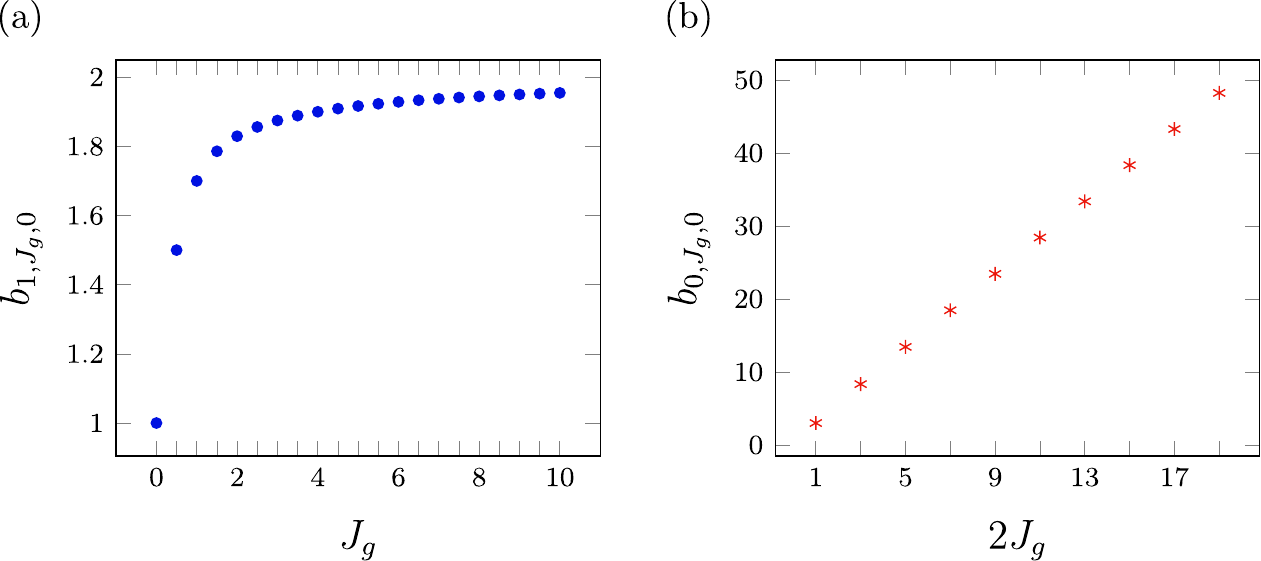}
\caption{(Color online) Values of (a) $b_{1, J_g, 0}$ and (b) $b_{0, J_g,0}$ as a function of $J_g$.}
\label{FigGao}
\end{centering}
\end{figure}

If all plane waves have the same frequency and the same pure polarization $q$, we have $\Omega = \sum_j \Omega_j$ and $\overline{R}_j$ simplifies to
\begin{equation} \label{SimplifyRj0}
\overline{R}_j = \frac{\Gamma}{2} s_j \frac{a_{\Delta J, J_g, q}}{b_{\Delta J, J_g, q} + s},
\end{equation}
with $s_j = \textrm{Re} [ (\Omega_j \Omega^* /\Gamma)/(\Gamma/2 - i \delta) ]$, $s = \sum_j s_j$ and where $\delta \equiv \delta_j, \forall j$. For $\Delta J = 1$, or $\Delta J = 0$ with $q=0$ and half-integer $J_g$, $a_{\Delta J, J_g, q} = 1$ and
\begin{equation}
b_{\Delta J, J_g, q} = \frac{\det{[A_{pp, +}^q]} + (-1)^{2J_g} \det{[A_{pp,-}^q]}}{\det{[A_{pp, +}^q]} - (-1)^{2J_g} \det{[A_{pp, -}^q]}} ,
\end{equation}
with $A_{pp,\pm}^q \equiv A_{pp} \pm (C_{pp})_q^q$. For $\Delta J = 0$ with $q \neq 0$ or integer $J_g$, $a_{\Delta J, J_g, q} = b_{\Delta J, J_g, q} = 0$. In particular, we have $b_{1,J_g,\pm 1}=1$. For $\Delta J = 1$, Eq.~(\ref{SimplifyRj0}) yields
\begin{equation} \label{fs1}
\overline{R}_j = \left\{ \begin{array}{ll} \displaystyle \frac{\Gamma}{2} \frac{s_j}{1+s} & \quad \textrm{for } q = \pm 1 , \vspace{0.1cm} \\
\displaystyle \frac{\Gamma}{2} \frac{s_j}{b_{1, J_g, 0} + s} & \quad \textrm{for } q=0 .
\end{array} \right.
\end{equation}
For $q = \pm 1$, in the periodic regime, the atom is pumped into the $\vert J_g , m_g = \pm J_g \rangle$ state from which it interacts only with the $\vert J_e, m_e = \pm J_e \rangle$ state through the laser radiation action. The atom then exactly behaves as a two-level system. For $q=0$, all populations are nonzero apart from $m_e = \pm J_e$ and the result is more subtle. For $\Delta J = 0$, Eq.~(\ref{SimplifyRj0}) yields $\overline{R}_j = 0$ for $q=\pm 1$ and
\begin{equation} \label{Rj0SamePolDeltaJ0q0}
\overline{R}_j = \left\{ \begin{array}{ll}
0 & \textrm{if $J_g$ is integer} , \\
\displaystyle \frac{\Gamma}{2} \frac{s_j}{b_{0, J_g, 0} + s} & \textrm{if $J_g$ is half-integer} ,
\end{array} \right.
\end{equation}
for $q=0$. For $q = \pm 1$, the atom is pumped into the $\vert J_g, m_g = \pm J_g \rangle$ state on which $\sigma^\pm$-radiation has no effect. For $q = 0$ and integer $J_g$, since the Clebsch-Gordan coefficient $\mathcal{C}_0^{(0)}$ is zero whatever $J_g$, the atom is pumped into the $\vert J_g, m_g = 0 \rangle$ state from which $\pi$-radiation has no effect. For $q=0$ and half-integer $J_g$, all populations are nonzero. We recall that our formalism does not apply for all lasers with same polarization in the $\Delta J = -1$ case [since $\det{(A_{\xi \xi}^{(0)} + B_{\xi \xi}^{(0, 0)})} = 0$]. Equations~(\ref{fs1}) and~(\ref{Rj0SamePolDeltaJ0q0}) perfectly reproduce the results of Ref.~\cite{Gao1993} that investigates those specific configurations. We show in Fig.~\ref{FigGao} the parameters $b_{1, J_g, 0}$ and $b_{0, J_g, 0}$ as a function of $J_g$.

\subsection{$N = 2$ case}

For $N = 2$ and $\omega_1 \neq \omega_2$, the system can be tackled in a continued fraction approach as in the two-level atom case~\cite{Podlecki2018}, though with a matrix formalism. The Fourier components $\mathbf{x}_\xi^{(n)}$ that obey the infinite system~(\ref{FinalSystFourier}) of equations subdivide in two decoupled groups: a first group of components of indices $n = k n_s$ ($k \in \mathbb{Z}$) and the group of all remaining components ($n \neq k n_s$), where $n_s = n_1 + n_2$, with $n_1$ and $n_2$ two positive coprime integers such that $n_2 \kappa_1 = n_1 \kappa_2$. The components of the first group get coupled between each other through the system
\begin{equation} \label{DegN2SystEBO}
\mathbf{x}_\xi^{(n)} + \mathcal{W}_{\xi \xi}^{(n, n_s)} \mathbf{x}_{\xi}^{(n+n_s)} + \mathcal{W}_{\xi \xi}^{(n, -n_s)} \mathbf{x}_\xi^{(n - n_s)} = \mathbf{d}_\xi \delta_{n, 0}.
\end{equation}
The components of the second group satisfy a homogeneous system and merely vanish. Therefore, the only \textit{a priori} nonzero $Q_{\xi \xi}^{(n)}$ matrices are for the $n$ indices of the first group, and similarly for the $R_j^{(n)}$ and $\mathbf{F}_j^{(n)}$ Fourier components (as for two-level atoms~\cite{Podlecki2018}). As long as one can define the matrices $\smash{Q_{\xi\xi}^{(n,m)}}$ that map $\mathbf{x}_\xi^{(n)}$ onto $\mathbf{x}_\xi^{(n+m)}, \forall m, n$: $Q_{\xi\xi}^{(n,m)} \mathbf{x}_\xi^{(n)} = \mathbf{x}_\xi^{(n+m)}$, Eq.~(\ref{DegN2SystEBO}) yields, $\forall n = k n_s \neq 0$, $Q_{\xi \xi}^{(n - n_s, n_s)} = - \mathcal{W}_{\xi \xi}^{(n, -n_s)}/( \mathbb{1}_\xi + \mathcal{W}_{\xi \xi}^{(n, n_s)} Q_{\xi \xi}^{(n, n_s)} )$ where $A/B \equiv B^{-1}A$. This implies $Q_{\xi \xi}^{(n_s)} = - \mathcal{W}_{\xi\xi}^{(n_s,-n_s)} / [ \mathbb{1}_\xi + \mathcal{K}_{k=1}^\infty (P_{\xi\xi,k}/\mathbb{1}_\xi) ]$, with $P_{\xi\xi,k} = - \mathcal{W}_{\xi\xi}^{(k n_s, n_s)} \mathcal{W}_{\xi\xi}^{((k+1) n_s, -n_s)}$ and $\mathcal{K}_{k=1}^\infty (P_{\xi\xi,k}/\mathbb{1}_\xi) \equiv P_{\xi\xi,1}/(\mathbb{1}_\xi + P_{\xi\xi,2}/(\mathbb{1}_\xi + P_{\xi\xi,3}/\ldots))$, provided all matrix inverses hold. For $n = kn_s$ with $k > 1$, Eq.~(\ref{DegN2SystEBO}) also yields the recurrence relation $Q_{\xi\xi}^{((k+1) n_s)} = - ( Q_{\xi\xi}^{(kn_s)} + {\mathcal{W}_{\xi \xi}^{(-kn_s,n_s)}}^* Q_{\xi\xi}^{((k-1)n_s)} )/\mathcal{W}_{\xi \xi}^{(kn_s,n_s)}$ that allows for the calculation of the remaining $Q_{\xi\xi}^{(kn_s)}$ along with all nonzero Fourier components $R_j^{(kn_s)}$ and $\mathbf{F}_j^{(kn_s)}$.

\section{Conclusion} \label{SecCcl}

In conclusion, we have extended the formalism of Ref.~\cite{Podlecki2018} to the multilevel atom case, where Zeeman sublevels and arbitrary light polarization are taken into account. In that context, we have provided a general standardized and exact analytical formalism for computing within the usual RWA the mechanical action experienced by a single multilevel atom lightened simultaneously by an arbitrary set of plane waves. By use of a Fourier expansion treatment, we provided an exact analytical expression of all Fourier components $\mathbf{F}_j^{(n)}$ describing the light forces in the periodic regime if unique. In particular, we extended the steady mean force expression~(\ref{F_KnownExpression}) into Eq.~(\ref{finalFj}), involving matrix quantities whose dimensions depend on the atomic structure. In addition, we highlighted some simplifications holding in specific regimes. The computation of the Fourier components related to the light forces relies on the solution of an algebraic system of equations and does not require numerical integration of the OBEs with time.

\acknowledgments
T.~Bastin acknowledges financial support of the Belgian F.R.S.-FNRS through IISN Grant 4.4512.08. The authors thank the Consortium des \'Equipements de Calcul Intensif (C\'ECI), funded by the Fonds de la Recherche Scientifique de Belgique (F.R.S.-FNRS) under Grant No. 2.5020.11 and by the Walloon Region, for providing the computational resources.

\appendix
\section{Reference frame rotation} \label{AppendixA}

The states $\vert J_g, m_g \rangle$ and $\vert J_e, m_e \rangle$ are common eigenstates of $\hat{\mathbf{J}}^2$ and $\hat{J}_z$, with $\hat{\mathbf{J}}$ the total angular momentum and $\hat{J}_z$  its $z$ component in the considered reference frame $S$. If $S$ is rotated according to Euler angles $\alpha$, $\beta$, $\gamma$ to a new configuration $\underline{S}$, the component $\hat{J}_z$ transforms to $\underline{\hat{J}}_z = \hat{R} (\alpha, \beta, \gamma) \hat{J}_z \hat{R} (\alpha, \beta, \gamma)^\dagger$, with the rotation operator $\hat{R} (\alpha, \beta, \gamma) = e^{-i \alpha \hat{J}_z/\hbar} e^{-i \beta \hat{J}_y/\hbar} e^{-i \gamma \hat{J}_z/\hbar}$~\cite{Sakurai1994}. $\hat{\mathbf{J}}^2$ remains unchanged and the common eigenstates of $\hat{\mathbf{J}}^2$ and $\underline{\hat{J}}_z$ read $\vert \underline{J_k , m_k} \rangle = \hat{R} (\alpha, \beta, \gamma) \vert J_k, m_k \rangle$ ($k=e,g$). The elements of the basis transformation matrix $\langle J_k , m_k \vert \underline{J_k , m_k^\prime} \rangle$ are given by the so-called Wigner functions $\mathcal{D}_{m_k, m_k^\prime}^{(J_k)} (\alpha, \beta, \gamma) = e^{- i (m_k \alpha + m_k^\prime \gamma)} d_{m_k, m_k^\prime}^{(J_k)} (\beta)$, with $d_{m, m^\prime}^{(J)} (\beta) = A_{m,m^\prime} (J) B_{m,m^\prime} (\beta) P^{(m^\prime-m, m^\prime+m)}_{J-m^\prime} (\cos \beta)$, where $A_{m,m^\prime} (J) = \sqrt{[(J+m^\prime)!(J-m^\prime)!]/[(J+m)!(J-m)!]}$, $B_{m,m^\prime} (\beta) = [\sin (\beta/2) ]^{m^\prime-m} [ \cos (\beta/2) ]^{m^\prime+m}$, and where the $P_n^{(a,b)} (z)$ are the Jacobi polynomials~\cite{Sakurai1994}. If $\rho (t)$ and $\underline{\rho} (t)$ denote the density matrices of the atomic state $\hat{\rho} (t)$ in the $\{ \vert J_k , m_k \rangle \}$ and $\{ \vert \underline{J_k , m_k} \rangle \}$ bases, respectively, we get
\begin{equation} \label{rhoTransformationLaw}
\underline{\rho} (t) = \mathcal{D} (\alpha, \beta, \gamma)^\dagger \rho (t) \mathcal{D} (\alpha, \beta, \gamma) ,
\end{equation}
with $\mathcal{D} (\alpha, \beta, \gamma) = \bigoplus_{k=e,g} \mathcal{D}^{(J_k)} (\alpha, \beta, \gamma)$, where $\mathcal{D}^{(J_k)} (\alpha, \beta, \gamma)$ is the unitary matrix of elements $\mathcal{D}_{m_k, m_k^\prime}^{(J_k)} (\alpha, \beta, \gamma)$. It follows that the associated OBE column vector $\mathbf{x} (t)$ transforms according to
\begin{equation}
\underline{\mathbf{x}} (t) = T (\alpha, \beta, \gamma) \mathbf{x} (t) ,
\label{xtlaw}
\end{equation}
with the transformation matrix
\begin{equation}
T (\alpha, \beta, \gamma) = \begin{pmatrix} T_{oo} (\alpha, \beta, \gamma) & 0 \\ 0 & T_{\xi\xi} (\alpha, \beta, \gamma) \end{pmatrix},
\end{equation}
where
\begin{equation}
T_{\xi\xi} (\alpha, \beta, \gamma) = \begin{pmatrix} T_{pp} (\beta) & T_{pZ} (\alpha, \beta) \\ T_{Zp} (\beta, \gamma) & T_{ZZ} (\alpha, \beta, \gamma) \end{pmatrix} ,
\end{equation}
with blocks $T_{oo} (\alpha, \beta, \gamma)$, $T_{pp} (\beta)$, $T_{pZ} (\alpha, \beta)$, $T_{Zp} (\beta, \gamma)$, and $T_{ZZ} (\alpha, \beta, \gamma)$ as explicitly detailed below. The OBEs~(\ref{MatrixOBE}) in the $\underline{S}$ reference frame then read
\begin{equation}
\dot{\underline{\mathbf{x}}} (t) = \underline{A} (t) \underline{\mathbf{x}} (t) + \underline{\mathbf{b}},
\end{equation}
with
\begin{equation} \label{A_TransformationLaw}
\underline{A} (t) = T (\alpha, \beta, \gamma) A (t) T (\alpha, \beta, \gamma)^{-1}
\end{equation}
and
\begin{equation} \label{b_TransformationLaw}
\underline{\mathbf{b}} = T (\alpha, \beta, \gamma) \mathbf{b} .
\end{equation}
Thanks to the orthogonality relations of the Clebsch-Gordan coefficients, to the orthogonality of the $d^{(1)} (\beta)$ matrices of elements $d^{(1)}_{m, m^\prime} (\beta)$, to the transformation law of the spherical components of any three-dimensional space vector $\boldsymbol{v}$, $(\underline{v}_1, \underline{v}_0, \underline{v}_{-1})^T = \mathcal{D}^{(1)}(\alpha, \beta, \gamma)^T (v_1,v_0,v_{-1})^T$, and to the identities $\mathcal{C}_{m_g}^{(q)} d_{m_e, m_g+q}^{(J_e)} (\beta) = \sum_{q^\prime} \mathcal{C}_{m_e-q^\prime}^{(q^\prime)} d_{m_e - q^\prime, m_g}^{(J_g)} (\beta) d_{q^\prime, q}^{(1)} (\beta)$ and $\mathcal{C}_{m_e - q}^{(q)} d_{m_e - q, m_g}^{(J_g)} (\beta) = \sum_{q^\prime} \mathcal{C}_{m_g}^{(q^\prime)} d_{m_e, m_g + q^\prime}^{(J_e)} (\beta) d_{q, q^\prime}^{(1)} (\beta)$ (see, e.g., Ref.~\cite{Biedenharn1984}), which imply, $\forall m_{k_1}, m_{k_2} = -J_k , \ldots , J_k$ ($k=e,g$), $\sum_q \mathcal{C}_{m_{g_1}}^{(q)} d_{m_{e_1}, m_{g_1}+q}^{(J_e)} (\beta) \mathcal{C}_{m_{g_2}}^{(q)} d_{m_{e_2}, m_{g_2}+q}^{(J_e)} (\beta) = \sum_q \mathcal{C}_{m_{e_1}-q}^{(q)} d_{m_{e_1} - q, m_{g_1}}^{(J_g)} (\beta) \mathcal{C}_{m_{e_2}-q}^{(q)} d_{m_{e_2}-q,m_{g_2}}^{(J_g)} (\beta)$, the explicit calculation of Eqs.~(\ref{A_TransformationLaw}) and~(\ref{b_TransformationLaw}) yields, as expected,
\begin{equation}
\underline{A} (t) = - \Gamma A_0 + \textrm{Im} ( \boldsymbol{\Omega} (t) \cdot \underline{\mathbf{e}}_C )
\end{equation}
and $\underline{\mathbf{b}} = \mathbf{b}$, where $\underline{\mathbf{e}}_C = \sum_q C^{(q)} \underline{\mathbf{e}}_q$, with $\underline{\mathbf{e}}_q$ the $\underline{S}$ lower- index spherical basis, such that $\boldsymbol{\Omega} (t) \cdot \underline{\mathbf{e}}_C = \sum_q \underline{\Omega}_q(t) C^{(q)}$.

Similarly, the transformation law $\underline{\mathbf{x}}_o (t) = T_{oo} (\alpha, \beta, \gamma) \mathbf{x}_o (t)$ [see Eq.~(\ref{xtlaw})] directly yields the standard contravariant transformation law $(\underline{\chi}_o^{(1)},\underline{\chi}_o^{(0)},\underline{\chi}_o^{(-1)})^T = \mathcal{D}^{(1)}(\alpha,\beta,\gamma)^\dagger (\chi_o^{(1)},\chi_o^{(0)},\chi_o^{(-1)})^T$ that proves the vectorial character of $\boldsymbol{\chi}_o(t)$ [see Eq.~(\ref{chijt})].

\subsection{The $T_{oo} (\alpha, \beta, \gamma)$ block}

In accordance with Eq.~(\ref{cohOpt}), the $T_{oo} (\alpha, \beta, \gamma)$ block is structured into vertically and horizontally ordered subblocks $T_{oo}^{(\Delta m,\Delta m^\prime)} (\alpha, \beta, \gamma)$, with respective indices $\Delta m$ and $\Delta m^\prime$ both ranging from $-(J_e+J_g)$ to $J_e+J_g$. The subblocks $T_{oo}^{(\Delta m,\Delta m^\prime)} (\alpha, \beta, \gamma)$ read $\tilde{T}_{oo}^{(\Delta m, \Delta m^\prime)} (\beta)\otimes U_+^{(\Delta m, \Delta m^\prime)} (\alpha, \gamma)$, with matrix elements $( \tilde{T}_{oo} ^{(\Delta m, \Delta m^\prime)} (\beta) )_{m, m^\prime} = d_{m^\prime, m}^{(J_g)} (\beta) d_{m^\prime + \Delta m^\prime, m+\Delta m}^{(J_e)} (\beta)$ ($m = m_-^{(\Delta m)}, \ldots, m_+^{(\Delta m)}$ and $m^\prime = m_-^{(\Delta m^\prime)}, \ldots, m_+^{(\Delta m^\prime)}$) and where
\begin{equation}
U_\pm^{(\Delta m, \Delta m^\prime)} (\alpha, \gamma) = \begin{pmatrix} c_\pm^{(\Delta m, \Delta m^\prime)}(\alpha, \gamma) & s_\pm^{(\Delta m, \Delta m^\prime)}(\alpha, \gamma) \\ \mp s_\pm^{(\Delta m, \Delta m^\prime)}(\alpha, \gamma) & \pm c_\pm^{(\Delta m, \Delta m^\prime)}(\alpha, \gamma) \end{pmatrix},
\end{equation}
with $c_\pm^{(\Delta m, \Delta m^\prime)}(\alpha, \gamma) = \cos (\Delta m^\prime \alpha \pm \Delta m \gamma)$ and $s_\pm^{(\Delta m, \Delta m^\prime)}(\alpha, \gamma) = \sin (\Delta m^\prime \alpha \pm \Delta m \gamma)$.

\subsection{The $T_{pp} (\beta)$ block}

The $T_{pp} (\beta)$ block is structured into 4 subblocks (one is zero) as
\begin{equation}
T_{pp} (\beta) = \begin{pmatrix} T_{p_ep_e} (\beta) & 0 \\ T_{p_gp_e} (\beta) & T_{p_gp_g} (\beta) \end{pmatrix},
\end{equation}
with subblock elements $( T_{p_ep_e} (\beta) )_{m_e, m_e^\prime} = ( d_{m_e^\prime, m_e}^{(J_e)} (\beta) )^2$, $( T_{p_gp_e} (\beta) )_{m_g, m_e^\prime} = - ( d_{-J_g, m_g}^{(J_g)} (\beta) )^2$, and $( T_{p_gp_g} (\beta) )_{m_g, m_g^\prime} = ( d_{m_g^\prime, m_g}^{(J_g)} (\beta) )^2 - ( d_{-J_g, m_g}^{(J_g)} (\beta) )^2$, where $m_k,m_k^\prime = - J_k + \delta_{k,g}, \ldots, J_k$ ($k=e,g$).

\subsection{The $T_{Zp} (\beta, \gamma)$ block}

The $T_{Zp} (\beta, \gamma)$ block is similarly structured into 4 subblocks (among which one is zero) as
\begin{equation}
T_{Zp} (\beta, \gamma) = \begin{pmatrix} T_{Z_ep_e} (\beta, \gamma) & 0 \\ T_{Z_gp_e} (\beta, \gamma) & T_{Z_gp_g} (\beta, \gamma) \end{pmatrix},
\end{equation}
where the subblocks $T_{Z_kp_l} (\beta, \gamma)$ ($k, l = g, e$) are themselves further divided [in accordance with Eq.~(\ref{cohZee})] into vertically ordered subsubblocks $T_{Z_kp_l}^{(\Delta m)} (\beta, \gamma)$ indexed with $\Delta m = 1, \ldots , 2J_k$. The subsubblocks $T_{Z_kp_l}^{(\Delta m)} (\beta, \gamma)$ read $\tilde{T}_{Z_kp_l}^{(\Delta m)} (\beta) \otimes (\cos (\Delta m \gamma), - \sin (\Delta m \gamma))^T$, with matrix elements $( \tilde{T}_{Z_ep_e}^{(\Delta m)} (\beta) )_{m_e,m_e^\prime} = d_{m_e^\prime, m_e}^{(J_e)} (\beta) d_{m_e^\prime, m_e + \Delta m}^{(J_e)} (\beta)$, $( \tilde{T}_{Z_gp_e}^{(\Delta m)} (\beta) )_{m_g,m_e^\prime} = - d_{-J_g, m_g}^{(J_g)} (\beta) d_{-J_g, m_g + \Delta m}^{(J_g)} (\beta)$, and $( \tilde{T}_{Z_gp_g}^{(\Delta m)} (\beta) )_{m_g,m_g^\prime} = d_{m_g^\prime, m_g}^{(J_g)} (\beta) d_{m_g^\prime, m_g + \Delta m}^{(J_g)} (\beta) - d_{-J_g, m_g}^{(J_g)} (\beta) d_{-J_g, m_g + \Delta m}^{(J_g)} (\beta)$, where $m_k =  -J_k, \ldots, J_k - \Delta m$ and $m_k^\prime = -J_k + \delta_{k, g}, \ldots, J_k$ ($k=e,g$).

\subsection{The $T_{pZ} (\alpha, \beta)$ block}

The $T_{pZ} (\alpha, \beta)$ block is structured into 2 diagonal subblocks as
\begin{equation}
T_{pZ} (\alpha, \beta) = \begin{pmatrix} T_{p_eZ_e} (\alpha, \beta) & 0 \\ 0 & T_{p_gZ_g} (\alpha, \beta) \end{pmatrix},
\end{equation}
where the subblocks $T_{p_kZ_k} (\alpha, \beta)$ ($k = e, g$) are themselves further divided [in accordance with Eq.~(\ref{cohZee})] into  horizontally ordered subsubblocks $T_{p_kZ_k}^{(\Delta m)} (\alpha, \beta)$ indexed with $\Delta m = 1, \ldots , 2J_k$. The subsubblocks $T_{p_kZ_k}^{(\Delta m)} (\alpha, \beta)$ read $\tilde{T}_{p_kZ_k}^{(\Delta m)} (\beta) \otimes (\cos (\Delta m \alpha), \sin (\Delta m \alpha) )$, with matrix elements $( \tilde{T}_{p_kZ_k}^{(\Delta m)} (\beta) )_{m_k,m_k^\prime} = 2 d_{m_k^\prime, m_k}^{(J_k)} (\beta) d_{m_k^\prime + \Delta m, m_k}^{(J_k)} (\beta)$, where $m_k =  -J_k + \delta_{k,g}, \ldots, J_k$ and $m_k^\prime = -J_k, \ldots, J_k - \Delta m$.

\subsection{The $T_{ZZ} (\alpha, \beta, \gamma)$ block}

The $T_{ZZ} (\alpha, \beta, \gamma)$ block is similarly structured into 2 diagonal subblocks as
\begin{equation}
T_{ZZ} (\alpha, \beta, \gamma) = \begin{pmatrix} T_{Z_eZ_e} (\alpha, \beta, \gamma) & 0 \\ 0 & T_{Z_gZ_g} (\alpha, \beta, \gamma) \end{pmatrix},
\end{equation}
where the subblocks $T_{Z_kZ_k} (\alpha, \beta, \gamma)$ ($k=e,g$) are themselves divided [in accordance with Eq.~(\ref{cohZee})] into vertically and horizontally subsubblocks $T_{Z_kZ_k}^{(\Delta m, \Delta m^\prime)} (\alpha, \beta, \gamma)$, with respective indices $\Delta m = 1, \ldots , 2J_k$ and $\Delta m^\prime = 1, \ldots , 2J_k^\prime$. The subsubblocks $T_{Z_kZ_k}^{(\Delta m, \Delta m^\prime)} (\alpha, \beta, \gamma)$ read $\sum_{\epsilon=\pm} \tilde{T}_{Z_kZ_k,\epsilon}^{(\Delta m, \Delta m^\prime)} (\beta) \otimes U_\epsilon^{(\Delta m, \Delta m^\prime)} (\alpha, \gamma)$, with matrix elements $( \tilde{T}_{Z_kZ_k,+}^{(\Delta m, \Delta m^\prime)} (\beta) )_{m_k,m_k^\prime} = d_{m_k^\prime, m_k}^{(J_k)} (\beta) d_{m_k^\prime + \Delta m^\prime, m_k + \Delta m}^{(J_k)} (\beta)$ and $( \tilde{T}_{Z_kZ_k,-}^{(\Delta m, \Delta m^\prime)} (\beta) )_{m_k,m_k^\prime} = d_{m_k^\prime + \Delta m^\prime, m_k}^{(J_k)} (\beta) d_{m_k^\prime, m_k + \Delta m}^{(J_k)} (\beta)$, where $m_k = -J_k, \ldots, J_k - \Delta m$ and $m_k^\prime = -J_k, \ldots, J_k - \Delta m^\prime$ ($k=e,g$).

\section{Explicit value of the $(C_{\xi\xi})^{q}_{q'}$ matrices} \label{AppendixB}

The matrices $(C_{\xi\xi})_{q^\prime}^q = - C_{\xi o}^{(q)} {C_{o \xi}^{(q^\prime)}}^*$ are structured into 4 blocks as
\begin{equation} \label{BlocksCxixiqqp}
(C_{\xi\xi})_{q^\prime}^q = \begin{pmatrix} (C_{pp})_{q^\prime}^q & (C_{pZ})_{q^\prime}^q \\ (C_{Zp})_{q^\prime}^q & (C_{ZZ})_{q^\prime}^q \end{pmatrix} ,
\end{equation}
with $(C_{rs})_{q^\prime}^q = - C_{ro}^{(q)} C_{os}^{(q^\prime)^{\scriptstyle{*}}}$ ($r,s = p, Z$). These blocks are themselves further divided into 4 subblocks as
\begin{equation}
(C_{rs})_{q^\prime}^q = \begin{pmatrix} (C_{r_e s_e})_{q^\prime}^q & (C_{r_e s_g})_{q^\prime}^q \\ (C_{r_g s_e})_{q^\prime}^q & (C_{r_g s_g})_{q^\prime}^q \end{pmatrix} ,
\end{equation}
where again $(C_{r_k s_l})_{q^\prime}^q = - C_{r_k o}^{(q)} C_{o s_l}^{(q^\prime)^{\scriptstyle{*}}}$ ($k, l = e,g$). These subblocks are detailed below.

\subsection{The $(C_{p_kp_l})_{q^\prime}^q$ subblocks}

For $k, l = e, g$, we have $(C_{p_kp_l})_{q^\prime}^q = ( \tilde{C}_{p_kp_l} )_q^q \delta_{q, q^\prime}$, with $( \tilde{C}_{p_kp_l} )_q^q = - \tilde{C}_{p_ko}^{(q)} \tilde{C}_{op_l}^{(q)}$. The $( \tilde{C}_{p_kp_l} )_q^q$ matrix elements are indexed with the two numbers $m = -J_k + \delta_{k,g} , \ldots , J_k$ and $m^\prime = -J_l + \delta_{l, g} , \ldots , J_l$. They are \textit{a priori} only nonzero if $m - n_k q \in \{ m_-^{(q)} , \ldots , m_+^{(q)} \}$, in which case they read explicitly $[ ( \tilde{C}_{p_kp_l} )_q^q ]_{m, m^\prime} = \tilde{n}_k ( \mathcal{C}_{m - n_k q}^{(q)} )^2 ( \delta_{m, - J_g + n_k q} + \tilde{n}_l \delta_{m^\prime, m + (n_l - n_k) q} ) / 2$. All $(C_{pp})_{q^\prime}^q $ matrix elements are real numbers.

\subsection{The $(C_{p_kZ_l})_{q^\prime}^q$ subblocks}

For $k, l = e, g$, we have
\begin{equation}
(C_{p_kZ_l})_{q^\prime}^q = \begin{pmatrix} ( \tilde{C}_{p_kZ_l}^{(1)} )_{q^\prime}^q & ( \tilde{C}_{p_kZ_l}^{(2)} )_{q^\prime}^q & 0 \end{pmatrix} .
\end{equation}
The $0$ block is of dimension $\dim{\mathbf{x}_{p_k}} \times \sum_{i=3}^{2J_l} \dim{\mathbf{x}_{Z_l}^{(i)}}$ and, for $j=1,2$,
\begin{equation} \label{CkZlqqp}
( \tilde{C}_{p_kZ_l}^{(j)} )_{q^\prime}^q = \left\{ \begin{array}{ll} 0_{\dim{\mathbf{x}_{p_k}} \times \dim{\mathbf{x}_{Z_l}^{(j)}}} & \textrm{if } \vert \Delta q \vert \neq j , \\
( \check{C}_{p_kZ_l}^{(j)} )_{q^\prime}^q \otimes ( 1 , \textrm{sgn} (\Delta q) i ) & \textrm{otherwise} , \end{array} \right.
\end{equation}
with $\Delta q \equiv q^\prime - q$ and $( \check{C}_{p_kZ_l}^{(j)} )_{q^\prime}^q = - \tilde{C}_{p_ko}^{(q)} \tilde{C}_{oZ_l, - \textrm{sgn} (\Delta q)}^{(q,\vert \Delta q \vert)}$ for $\Delta q \neq 0$ (see Section~\ref{SecForce}). Hence, $(C_{p_kZ_l})_q^q = 0$, $\forall q$. The $( \check{C}_{p_kZ_l}^{(j)} )_{q^\prime}^q$ matrix elements are indexed with the two numbers $m = -J_k + \delta_{k, g} , \ldots , J_k$ and $m^\prime = -J_l , \ldots , J_l - \vert \Delta q \vert$. They are \textit{a priori} only nonzero if $m - n_k q \in \{ m_-^{(q)} , \ldots , m_+^{(q)} \}$, in which case they read explicitly $[ ( \check{C}_{p_kZ_l}^{(j)} )_{q^\prime}^q ]_{m, m^\prime} = (\tilde{n}_k \tilde{n}_l \mathcal{C}_{m - n_k q}^{(q)} \mathcal{C}_{m - n_k q + (n_l - 1) \Delta q}^{(q^\prime)} \delta_{m^\prime, m + (\tilde{n}_l q^\prime - \tilde{n}_k q -j)/2}) / 2$.

\subsection{The $(C_{Z_kp_l})_{q^\prime}^q$ subblocks}

For $k, l = e, g$, we have
\begin{equation}
(C_{Z_kp_l})_{q^\prime}^q = \begin{pmatrix} ( \tilde{C}_{Z_kp_l}^{(1)} )_{q^\prime}^q \\ ( \tilde{C}_{Z_kp_l}^{(2)} )_{q^\prime}^q \\ 0 \end{pmatrix} .
\end{equation}
The $0$ block is of dimension $\sum_{i=3}^{2J_k} \dim{\mathbf{x}_{Z_k}^{(i)}} \times \dim{\mathbf{x}_{p_l}}$ and, for $j=1,2$,
\begin{equation} \label{CkZlqqpAndCZklqqp}
( \tilde{C}_{Z_kp_l}^{(j)} )_{q^\prime}^q = \left\{ \begin{array}{ll} 0_{\dim{\mathbf{x}_{Z_k}^{(j)}} \times \dim{\mathbf{x}_{p_l}}} & \textrm{if } \vert \Delta q \vert \neq j , \\
( \check{C}_{Z_kp_l}^{(j)} )_{q^\prime}^q \otimes ( 1 , \textrm{sgn} (\Delta q) i )^T & \textrm{otherwise} , \end{array} \right.
\end{equation}
with $( \check{C}_{Z_kp_l}^{(j)} )_{q^\prime}^q = - \tilde{C}_{Z_k o, \textrm{sgn} (\Delta q)}^{(\vert \Delta q \vert, q^\prime)} \tilde{C}_{op_l}^{(q^\prime)}$ for $\Delta q \neq 0$, where $\tilde{C}_{Z_ko, \epsilon}^{(\vert \Delta q \vert  , q^\prime)} = - \tilde{C}_{oZ_k, \epsilon}^{(q^\prime, \vert \Delta q \vert )^{\scriptstyle{T}}}$ for $\epsilon = \pm 1$ (see Section~\ref{SecForce}). Hence, $(C_{Z_kp_l})_q^q = 0  , \forall q$. The $( \check{C}_{Z_kp_l}^{(j)} )_{q^\prime}^q$ matrix elements are indexed with the two numbers $m = -J_k , \ldots , J_k - \vert \Delta q \vert$ and $m^\prime = -J_l + \delta_{l, g} , \ldots , J_l$. They are \textit{a priori} only nonzero if $m -  n_k q - [1-\textrm{sgn} (\Delta q)] \Delta q / 2 \in \{ m_-^{(q^\prime)} , \ldots , m_+^{(q^\prime)} \}$, in which case they read explicitly $[ ( \check{C}_{Z_kp_l}^{(j)} )_{q^\prime}^q ]_{m, m^\prime} = (\tilde{n}_k \mathcal{C}_{-J_g}^{(q^\prime)} \mathcal{C}_{-J_g-(n_k-1)\Delta q}^{(q)} \delta_{m,-J_g + (q^\prime + \tilde{n}_k q - j)/2} + \tilde{n}_k \tilde{n}_l \mathcal{C}_{m^\prime - n_l q^\prime}^{(q^\prime)} \penalty 0 \mathcal{C}_{m^\prime - n_l q^\prime - (n_k-1) \Delta q}^{(q)} \delta_{m^\prime, m +(\tilde{n}_l q^\prime - \tilde{n}_k q + j)/2}) /4$. We note that $(C_{p_kZ_l})_{q^\prime}^q \neq   - [ (C_{Z_lp_k})_{q^\prime}^q ]^T$.

\subsection{The $(C_{Z_kZ_l})_{q^\prime}^q$ subblocks}

For $k, l = e, g$, we have
\begin{equation} \label{GeneralCZkZlExpression}
\begin{split}
(C_{Z_kZ_l})_{q^\prime}^q = & \sum\limits_{\epsilon = \pm 1} ( \tilde{C}_{Z_kZ_l, \epsilon} )_{q^\prime}^q \otimes \begin{pmatrix} 1 & - \epsilon i \\ \epsilon i & 1 \end{pmatrix} \\
& + ( \tilde{C}_{Z_kZ_l} )_q^q \otimes \begin{pmatrix} 1 & - q i \\ - q i & - 1 \end{pmatrix} \delta_{q^\prime, -q},
\end{split}
\end{equation}
with $( \tilde{C}_{Z_kZ_l, \epsilon} )_{q^\prime}^q$ and $( \tilde{C}_{Z_kZ_l} )_q^q$ as described below.

If $\textrm{diag}_p (X_1 , \ldots , X_n)$ denotes the rectangular matrix whose elements are matrix blocks, with $X_1 , \ldots , X_n$ the only nonzero such elements exactly located on the $\smash{p^{\textrm{th}}}$ superdiagonal (or subdiagonal if $p<0$) of the rectangular matrix, then we have
\begin{equation} \label{ExprTildeCZkZlqqp}
( \tilde{C}_{Z_kZ_l, \epsilon} )_{q^\prime}^q = \textrm{diag}_{- \epsilon \Delta q} \left( (\tilde{C}_{Z_kZ_l, \epsilon}^{(\Delta m_-)})_{q^\prime}^q , \ldots , (\tilde{C}_{Z_kZ_l, \epsilon}^{(\Delta m_+)})_{q^\prime}^q \right) ,
\end{equation}
where $\Delta m_- = \max{[1, 1 + \epsilon \Delta q]}$, $\Delta m_+ = \Delta m_- + 2 \min{(J_e, J_g)} - 1$, and, for $\Delta m = \Delta m_- , \ldots , \Delta m_+$, $( \tilde{C}_{Z_kZ_l,\epsilon}^{(\Delta m)} )_{q^\prime}^q$ is a matrix block of dimension $(\dim{\mathbf{x}_{Z_k}^{(\Delta m)}}/2) \times (\dim{\mathbf{x}_{Z_l}^{(\Delta m - \epsilon \Delta q)}}/2)$. These blocks are \textit{a priori} only nonzero for $\Delta m + \epsilon q \leq J_e+J_g$, in which case they identify to $- \tilde{C}_{Z_k o,\epsilon}^{(\Delta m , q + \epsilon \Delta m)} \tilde{C}_{oZ_l, \epsilon}^{(q + \epsilon \Delta m, \Delta m - \epsilon \Delta q)}$, with matrix elements $[ (\tilde{C}_{Z_kZ_l, \epsilon}^{(\Delta m)})_{q^\prime}^q ]_{m, m^\prime} = (\tilde{n}_k \tilde{n}_l/4) \penalty 0 \mathcal{C}_{m - (1+\tilde{n}_k) q/2 + (1 -\epsilon \tilde{n}_k) \Delta m/2}^{(q)} \penalty 0 \mathcal{C}_{m + (1-\epsilon\tilde{n}_l)\Delta m/2 - (\tilde{n}_k q - \tilde{n}_l \Delta q + q^\prime)/2}^{(q^\prime)} \penalty 0 \delta_{m^\prime, \mu}$, where $\mu = m + (\tilde{n}_l q^\prime - \tilde{n}_k q + \epsilon \Delta q) / 2$, $m = -J_k , \ldots , J_k - \Delta m$, and $m^\prime = -J_l , \ldots , J_l - \Delta m + \epsilon \Delta q$.

We also have
\begin{equation} \label{ExprTildeCZkZlq}
( \tilde{C}_{Z_kZ_l} )_q^q = \left\{ \begin{array}{ll}
0_{(\dim{\mathbf{x}_{Z_k}}/2) \times (\dim{\mathbf{x}_{Z_l}}/2)} & \quad \textrm{if } q = 0 , \vspace{0.2cm} \\
\begin{pmatrix}
( \check{C}_{Z_kZ_l} )_q^q & 0 \\ 0 & 0_{d_k \times d_l}
\end{pmatrix} & \quad \textrm{otherwise,} \end{array} \right.
\end{equation}
with, for $k = e, g$, $d_k \equiv \sum_{i=2}^{2J_k} \dim{\mathbf{x}_{Z_k}^{(i)}}/2 = J_k (2 J_k - 1)$ and $( \check{C}_{Z_kZ_l} )_q^q = - \tilde{C}_{Z_k o,-q}^{(1,0)} \tilde{C}_{oZ_l,q}^{(0,1)}$. The $( \check{C}_{Z_kZ_l} )_q^q$ matrix elements are indexed with the two numbers $m = -J_k, \ldots , J_k-1$ and $m^\prime = -J_l , \ldots, J_l - 1$. They read explicitly $[ ( \check{C}_{Z_kZ_l} )_q^q ]_{m, m^\prime} = (\tilde{n}_k \tilde{n}_l / 4) \mathcal{C}_{m + (1-q)/2}^{(q)} \mathcal{C}_{m^\prime + (1+q)/2}^{(-q)} \delta_{m^\prime, m-(\tilde{n}_k+\tilde{n}_l)q/2}$.

The $(C_{Z_kZ_l})_{q^\prime}^q$ subblocks are such that $(C_{ZZ})_{q^\prime}^q = [(C_{ZZ})^{q^\prime}_q ]^\dagger$ and $(C_{Z_gZ_e})_{q^\prime}^q = [ (C_{Z_eZ_g})_{q^\prime}^q ]^T$, $\forall q, q^\prime$. In addition, $\sum_q (C_{ZZ})_q^q$ is a real matrix.

\end{document}